\RequirePackage{fix-cm}
\documentclass[smallextended]{svjour3}       
\smartqed  
\usepackage{graphicx}
\usepackage{algorithmic}
\usepackage{algorithm}
\usepackage{multicol}
\usepackage{lipsum}
\usepackage{amsmath}
\usepackage{booktabs}
\usepackage{mathptmx}      
\usepackage{appendix}

\begin{document}

\title{A Unitary Operator Construction Solution Based on Pauli Group for Maximal Dense Coding 
}


\author{Wenjie Liu \and
        Junxiu Chen \and
        Wenbin Yu \and
        Zhihao Liu \and
        Hanwu Chen \and
}


\institute{W. Liu \at
              Engineering Research Center of Digital Forensics, Ministry of Education \\
              School of Computer and Software, Nanjing University of Information Science and Technology, Nanjing 210044, China \\
              \email{wenjiel@163.com}           
           \and
           J. Chen and W. Yu \at
              School of Computer and Software, Nanjing University of Information Science and Technology, Nanjing 210044, China
           \and
           Z. Liu and H. Chen \at
              School of Computer Science and Engineering, Southeast University, Nanjing 211189, China
}

\date{Received: date / Accepted: date}

\maketitle

\begin{abstract}
Quantum dense coding plays an important role in quantum cryptography communication, and
how to select a set of appropriate unitary operators to encode message is the primary work in the design of quantum communication protocols. Shukla et al. proposed a preliminary method for unitary operator construction based on Pauli group under multiplication, which is used for dense coding in quantum dialogue. However, this method lacks feasible steps or conditions,
and cannot construct all the possible unitary operator sets. In this study, a feasible solution of constructing unitary operator sets for quantum maximal dense
coding is proposed, which aims to use minimum qubits to maximally encode a class of $t$-qubit symmetric states. These states have an even number of superposition items, and there is at least one set of $\left\lceil {{t \over 2}} \right\rceil $ qubits whose superposition items are orthogonal to each other. Firstly, we propose the procedure and the corresponding algorithm for constructing ${2^{t}}$-order multiplicative modified generalized Pauli subgroups (multiplicative MGP subgroups). Then, two conditions for $t$-qubit symmetric states are given to select appropriate unitary operator sets from
the above subgroups. Finally, we take 3-qubit GHZ, 4-qubit W, 4-qubit cluster and 5-qubit cluster states as examples, and demonstrate how to find all unitary operator sets for maximal dense coding through our construction solution, which shows that our solution is feasible and convenient.
\end{abstract}

\section{Introduction}
\begin{figure*}[!t]
  \includegraphics[width=4.5in]{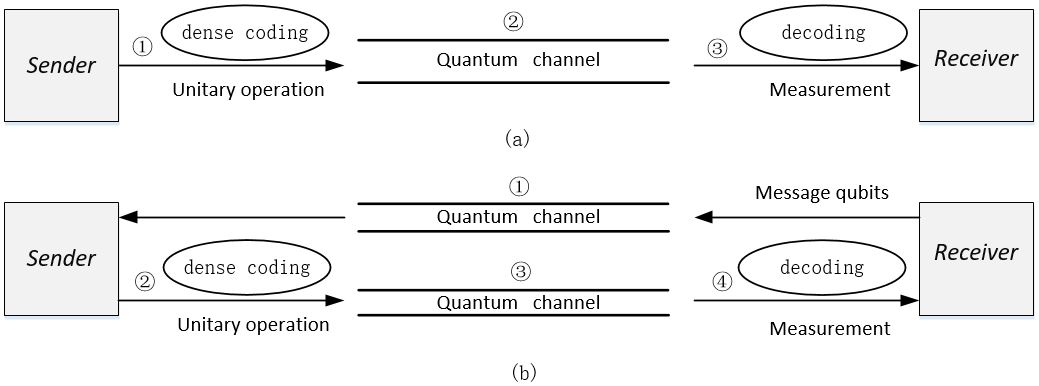}\\
  \caption{Two types of quantum communication process based on dense coding: (a) is one-way communication and (b) is two-way communication. In the one-way communication, the sender firstly encodes the classical information into quantum states using dense coding, and transmits them to the receiver through the quantum channel. Finally, the receiver gets the transmitted states and measures them to get the classical information (i.e., quantum decoding). In the two-way communication, the receiver firstly send message qubits (which are used for massage encoding) to the sender, and the remaining steps are the same as the one-way communication process.}
  \label{fig04}
\end{figure*}

With the development of quantum technology, quantum cryptography has become a hot topic, which has attracted more and more attention in the field of cryptography and physics. Many kinds of quantum cryptography protocols have been proposed, including quantum key distribution (QKD)~\cite{Bennett2014Public,Paul2013Experimental,Vlachou2018Quantum}, quantum secret sharing (QSS)~\cite{Hillery1999Quantum,Cleve1999How,Mashhadi2019General,Gottesman2000Theory}, quantum secure direct communication (QSDC)~\cite{Liu2008Efficient,Yin2018Quantum,Hu2016Experimental,Liu2009An}, quantum key agreement (QKA)~\cite{Huang2017Efficient,Liu2018AnEfficient,Lin2019Cryptanalysis,Cao2017Multiparty}, and quantum machine learning~\cite{Sheng2017Distributed,Liu2018Quantum,Biamonte2017Quantum,Liu2019AUnitary} has recently become a research hotspot. Quantum dense coding~\cite{Mattle1996Dense,Li2007Complete,Ambainis2002Dense} is a frequently used method in quantum information theory, and it plays an important role in quantum cryptography communication. In general, communication process based on dense coding can be divided into two categories: one-way and two-way communication, which can be illustrated in in Fig.~\ref{fig04}.

In quantum information theory, a qubit can be utilized to transmit more than one bits of classic information, which is called dense coding. However, maximal dense coding is a special type of dense coding, which require a qubit to transmit two
bits of classical information, to be more accurate, in a dense coding scheme, a $t$-qubit preshared entanglement can be traded as at most ${2^t}$ classical bits given that there are an error-free qubit channel and an error-free pre-shared entanglement~\cite{Hsieh2010Trading,Khalighi2014Survey}. The main focus of this paper is about the maximal dense coding. As we all know, how to select a set of appropriate unitary operators to encode message is the primary work in the design of quantum communication protocols.
Suppose the initial state is $t$-qubit state $\left| {{\phi _0}} \right\rangle $, in order
to implement maximal dense coding, we need ${2^t}$ unitary operators $\left\{ {{U_0},{U_1},{U_2}, \cdots ,{U_{{2^t} - 1}}} \right\}$
to operate on $\left| {{\phi _0}} \right\rangle $, and then get ${2^t}$ mutually orthogonal state vectors $\left\{ {\left| {{\phi _0}} \right\rangle ,\left|
{{\phi _{\rm{1}}}} \right\rangle , \cdots ,\left| {{\phi _i}} \right\rangle , \cdots ,\left| {{\phi _{{2^t} - 1}}} \right\rangle }
 \right\}$, where $\left| {{\phi _i}} \right\rangle = {U_i}\left| {{\phi _0}} \right\rangle $
 ($0 \le i \le {2^t} - 1$).

However, how to construct an appropriate unitary operator set
$\left\{ {{U_0},{U_1},{U_2}, \cdots ,{U_{{2^t} - 1}}} \right\}$ to encode message is not an easy work. Taking 3-qubit GHZ
state as an example, we require an 8-order unitary operator group to operate $\left\lceil {{{3} \over 2}} \right\rceil
{\rm{ = }}2$ qubits, which means we need select eight operators from
${G_2} = \left\{ {I,X,Y,Z} \right\} \otimes \left\{ {I,X,Y,Z} \right\}$, and there are
$C_{16}^8=12870$ possible combinations. Obviously, not all combinations are multiplicative groups, and it will be a hard work if we check them one by one. At present, its construction mostly
depends on the non-systematic search. Meanwhile, as the number of qubits increases, this experience-based method becomes more and
more difficult. So, it is of both theoretical and practical importance to find out an effective solution of constructing unitary operator sets.

In 2013, Shukla et al.~\cite{Shukla2013On} proposed a preliminary method for unitary operator construction based on the multiplication group, which is used for dense coding in quantum dialogue. They firstly construct a few subgroups of $G_n$ ($n$-fold tensor products of Pauli matrices, $n$ indicates the number of operated qubits) through the subgroups of $G_1^{'}$ (i.e., $\left\{ {I,X,Y,Z} \right\},\left\{ {I,X} \right\},\left\{ {I,Y} \right\},\left\{ {I,Z} \right\}$) , and then select appropriate unitary operator sets for a specific quantum state according to mutually-orthogonal-state distinguishable principle. However, this method is not feasible due to lack of detailed steps or conditions, what's more, it cannot construct all the possible unitary operator sets. In order to solve these problems, a feasible solution of constructing unitary operator sets for quantum maximal dense coding is proposed, which uses minimum qubits to encode a class of $t$-qubit symmetric states. These states have two constraints: (1) The number of superposition items of the quantum state must be even. (2) There is at least one set of $\left\lceil {{t \over 2}} \right\rceil $ qubits whose superposition items are orthogonal to each other.

The remaining parts of the paper are organized as follows. In Section 2, some preliminaries, i.e., the modified generalized Pauli
group, the Sylow theorem and quantum distinguishability principle, are introduced. In Section 3, Shukla et al.'s preliminary method for unitary operator construction is briefly reviewed.
And our solution of constructing unitary operator sets for quantum  maximal dense
coding is detailedly introduced in Section 4, which consists of two phases: constructing
multiplicative MGP subgroups, and selecting appropriate unitary operator sets.
In Section 4.1, the procedure and corresponding algorithm for constructing ${2^t}$-order multiplicative MGP subgroups are described
 elaborately.
 After that, two conditions for quantum symmetric states are introduced concretely in Section 4.2.
Taking 3-qubit GHZ state, 4-qubit W, 4-qubit cluster and 5-qubit cluster states as quantum resource,
we demonstrate that how to find all appropriate unitary operator sets through the above solution. Finally,
Section 5 is dedicated for conclusion.
\section{Preliminaries}
\subsection{The modified generalized Pauli group}
As we know, Pauli group is defined as ${P_1}$ = $\{  \pm {\sigma_i}, \pm i{\sigma _i}, \pm {\sigma _x}, \pm i{\sigma _x}, \pm {\sigma _y}, \pm i{\sigma _y}, \pm {\sigma _z}, \pm i{\sigma _z}\}$. Here, the inclusion of $ \pm 1$ and $ \pm i$ ensures that ${P_1}$ is closed under normal multiplication, but in quantum mechanics, the effects of $ \pm {\sigma _i}$ and $ \pm i{\sigma _i}$ on a quantum state are same. So we ignore the global phase from the product of matrices to redefine the multiplication operator for
two elements of the group. Since the Hilbert space of $t$-qubit system is ${C^{{2^{t}}}}$, then we need ${2^{t}}$ unitary operators to get ${2^{t}}$ mutually orthogonal state vectors. In this paper, the unitary operators are formed by combining Pauli operators (more precisely by combining $\left\{ {{\sigma _i},{\sigma _x},i{\sigma _y},{\sigma _z}} \right\}$). It is straightforward that $\left\{ {{\sigma _i},{\sigma _x},i{\sigma _y},{\sigma _z}} \right\}$ forms a group under multiplication (without global phase), which is called modified Pauli group. Thus we obtain a modified Pauli group ${G_1} = \left\{ {{\sigma _i},{\sigma _x},i{\sigma _y},{\sigma _z}} \right\}$ = $\left\{ {I,X,Y,Z} \right\}$ where the effect of $I,X,Y,Z$ on $\left| 0 \right\rangle$ and
$\left| 1 \right\rangle$ are described as Eq.~\ref{eqn:11},

\begin{equation}
\label{eqn:11}
\begin{array}{*{20}{l}}

&\left\{ \begin{matrix}
   {I\left| 0 \right\rangle  \to \left| 0 \right\rangle }  \cr
   {I\left| 1 \right\rangle  \to \left| 1 \right\rangle }  \cr
 \end{matrix}
 \right. ,\;\;\;\;\;\left\{ \begin{matrix}
   {X\left| 0 \right\rangle  \to \left| 1 \right\rangle }  \cr
   {X\left| 1 \right\rangle  \to \left| 0 \right\rangle }  \cr
 \end{matrix}
 \right., \cr
 \\
&\left\{ \begin{matrix}
   {Y\left| 0 \right\rangle  \to  - \left| 1 \right\rangle }  \cr
   {Y\left| 1 \right\rangle  \to \left| 0 \right\rangle }  \cr
 \end{matrix}
 \right.,\;\left\{ \begin{matrix}
   {Z\left| 0 \right\rangle  \to \left| 0 \right\rangle }  \cr
   {Z\left| 1 \right\rangle  \to  - \left| 1 \right\rangle }  \cr
 \end{matrix}
 \right.,
 \end{array}
\end{equation}
here, $Y = ZX$. Now we may define the  modified generalized Pauli group (MGP group) ${G_n}$ as one whose elements are all $n$-fold tensor products of Pauli matrices, i.e., ${G_n} = {G_1} \otimes {G_1} \otimes  \cdots  \otimes {G_1}$. Take ${G_2}$ as an example,
\begin{equation}
\label{eqn:01}
\begin{array}{*{20}{l}}
  & {G_2}\; = {G_1} \otimes {G_1} = \left\{ {I,X,Y,Z} \right\} \otimes \left\{ {I,X,Y,Z} \right\}  \cr
  & \;\;\;\;\;\;\; = \{ I \otimes I,I \otimes X,I \otimes Y,I \otimes Z,X \otimes I,X \otimes X,X \otimes Y,X \otimes Z, \cr
  & \;\;\;\;\;\;\;\;\;\;\; Y \otimes I,Y \otimes X,Y \otimes Y,Y \otimes Z,Z \otimes I,Z \otimes X,Z \otimes Y,Z \otimes Z\}  \cr
\end{array}
\end{equation}
is a group of order 16. In general, ${G_n} = {G_1}^{ \otimes n}$ is a group of order ${2^{2n}} = {\rm{ }}{4^n}$.

\subsection{Sylow theorem}

\paragraph{Sylow theorem:}
\emph{If G is a group of order ${p^k}m$, with p prime, and (p, m) = 1. For every prime factor p, there exists a Sylow p-subgroup of G and indeed the order of a Sylow p-subgroup is ${p^i}$ $\left( {i \le k} \right)$. and the Sylow p-subgroups of a group (for a given prime p) are conjugate to each other.}

In this paper, we utilize a ${4^n}$-order ${G_n}$. For ${4^n}$-order ${G_n}$, $p = 2, m = 1, k = 2n$ and in general ${G_n}$ has Sylow 2-subgroups of order $2,{\rm{ }}4,{\rm{ }}8,{\rm{ }}.{\rm{ }}.{\rm{ }}.{\rm{ }},{2^{k}}$. Thus ${G_2}$ has subgroups of order 16, 8, 4, 2, 1. Obviously the largest subgroup is the group ${G_n}$ itself and it has only one subgroup of order ${2^{k}}$. But we are concerned about subgroups of order ${2^i}$ where $0 \le i \le k$ in general and specially about subgroups of order ${2^{k - 1}}$. To be specific, giving $t$-qubit initial states, we utilize ${2^t}$ unitary operators on initial state to get ${2^t}$ mutually orthogonal states. So we must operate at least $\left\lceil {{t \over 2}} \right\rceil $ qubit. Thus the group ${G_n}$ must satisfy that $\left\lceil {{t \over 2}} \right\rceil  \le n \le t$ ($n$ represents the number of operated qubits).

\subsection{Quantum distinguishability principle}

The indistinguishability of non-orthogonal quantum states is at the heart of quantum
computation and quantum information. It is the essence of our assertion that a quantum
state contains hidden information that is not accessible to measurement, and thus
plays a key role in quantum algorithms and quantum cryptography~\cite{Nielson2000Quantum}.
The indistinguishability of qubits can be explained as follows: if two qubits $\left| \varphi  \right\rangle$ and
$\left| \phi  \right\rangle $ are satisfied
\begin{equation}
\label{eqn:12}
{\left\| {\left\langle {\varphi \left| \phi  \right\rangle } \right.} \right\|^2}{\rm{ = }}\cos \theta,
\end{equation}
here $\theta$ is the angle between two non-orthogonal qubits, and $0 < \theta  < {\pi  \over 2}$, $\cos \theta  \ne 0$,
then the two qubits are indistinguishable.
The indistinguishability of two qubits means that no precise results can be obtained for any operation or measurement. For example,
operating on either of the two qubits will inevitably produce some incorrect results with a certain probability.
Eq. \ref{eqn:12} also define indistinguishability degree as below,
\begin{equation}
\label{eqn:13}
{D = \left\| {\left\langle {\varphi \left| \phi  \right\rangle } \right.} \right\|^2}{\rm{ = }}\cos \theta,
\end{equation}
here $0 \le D \le 1$. Obviously, $D=1$ represents that these two qubits are absolutely indistinguishable;
$D=0$ represents that these two qubits are absolutely distinguishable. That is to say, those mutually orthogonal states are distinguishable~\cite{Nielson2000Quantum} (we call it as the mutually-orthogonal-state distinguishable principle in this paper).

\emph{Distinguishability}
is most easily understood using the metaphor of a game involving two parties, Alice and Bob.
Suppose a fixed set of states
$\left\{ {\left| {{\varphi _1}} \right\rangle ,\left| {{\varphi _2}} \right\rangle ,
\cdots ,\left| {{\varphi _n}} \right\rangle } \right\}$ is given, Alice chooses a
state $\left| {{\varphi _i}} \right\rangle \left( {1 \le i \le n} \right)$ from the set. She gives the state $\left| {{\varphi _i}} \right\rangle$ to Bob, whose task it is to identify the index $i$ of the state Alice has given him. Suppose the states $\left| {{\varphi _i}} \right\rangle$ are orthonormal. Then Bob can do a quantum measurement
to distinguish these states, using the following procedure. Define measurement operators
${M_i} \equiv \left| {{\varphi _i}} \right\rangle \left\langle {{\varphi _i}} \right|$,
one for each possible index i, and an additional measurement operator
$M_0$ defined as the positive square root of the positive operator $I - \sum\nolimits_{i \ne 0}^{} {\left| {{\varphi _i}}
\right\rangle \left\langle {{\varphi _i}} \right|} $.
These operators satisfy the completeness relation, and if the state $\left| {{\varphi _i}} \right\rangle$ is prepared
then the probability of obtaining measurement
outcome $i$ is
$p\left( i \right) = \left\langle {{\varphi _i}} \right|{M_i}\left| {{\varphi _i}} \right\rangle  = 1$,
so the result $i$ occurs with certainty. Thus, it is possible to
reliably distinguish the orthonormal states $\left| {{\varphi _i}} \right\rangle$.

\section{Review on Shukla et al.'s preliminary method for unitary operator construction}

In 2013, Shukla et al.~\cite{Shukla2013On} proposed a preliminary method for unitary operator
construction based on the group under multiplication, which is used for maximal dense coding in quantum
dialogue. For a certain $t$-qubit quantum state, these operators will be at least $\left\lceil {{t \over 2}} \right\rceil $-qubit operators (according to Sylow theorem in Section 2.2). So, for
3-qubit states, we need the 8-order subgroups of $G_2$  to complete the maximal dense coding.
Since each Pauli gate is self inverse, so $\left\{ {I,X} \right\},\left\{ {I,Y} \right\},\left\{ {I,Z} \right\}$ are subgroups
of $G_1$ consequently. The following are 8-order subgroups of $G_2$:

  $\begin{array}{l}G_2^1\left( 8 \right){\rm{ = }}{G_1} \otimes \left\{ {I,X} \right\}\\
\;\;\;\;\;\;\;\;\;\; =\left\{ I \otimes I,I \otimes X,X \otimes I,X \otimes X,Y \otimes I,Y \otimes X,Z \otimes I,Z \otimes X \right\}
\end{array}$

  $\begin{array}{l}G_2^2\left( 8 \right){\rm{ = }}{G_1} \otimes \left\{ {I,Y} \right\}\\
\;\;\;\;\;\;\;\;\;\; =\left\{ I \otimes I,I \otimes Y,X \otimes I,X \otimes Y,Y \otimes I,Y \otimes Y,Z \otimes I,Z \otimes Y \right\}
\end{array}$

  $\begin{array}{l}G_2^3\left( 8 \right){\rm{ = }}{G_1} \otimes \left\{ {I,Z} \right\}\\
\;\;\;\;\;\;\;\;\;\; =\left\{ I \otimes I,I \otimes Z,X \otimes I,X \otimes Z,Y \otimes I,Y \otimes Z,Z \otimes I,Z \otimes Z \right\}
\end{array}$

  $\begin{array}{l}G_2^4\left( 8 \right){\rm{ = }}\left\{ {I,X} \right\}\otimes {G_1}\\
\;\;\;\;\;\;\;\;\;\; =\left\{ I \otimes I,I \otimes X,X \otimes I,X \otimes X,Y \otimes I,Y \otimes X,Z \otimes I,Z \otimes X \right\}
\end{array}$

  $\begin{array}{l}G_2^5\left( 8 \right){\rm{ = }}\left\{ {I,Y} \right\}\otimes {G_1}\\
\;\;\;\;\;\;\;\;\;\; =\left\{ I \otimes I,Y \otimes I,I \otimes X,Y \otimes X,I \otimes Y,Y \otimes Y,I \otimes Z,Y \otimes Z \right\}
\end{array}$

  $\begin{array}{l}G_2^6\left( 8 \right){\rm{ = }}\left\{ {I,Z} \right\}\otimes {G_1}\\
  \;\;\;\;\;\;\;\;\;\; =\left\{ I \otimes I,Z \otimes I,I \otimes X,Z \otimes X,I \otimes Y,Z \otimes Y,I \otimes Z,Z \otimes Z \right\}.
\end{array}$
\\
where $G_n^j\left( m \right)$ denotes $j$th subgroup of $m$-order (${m < 4^n}$) of the
group $G_n$ whose order is $4^n$. Similarly, it can construct $3n$ subgroups
of order ${{\rm{2}}^{2n - 1}}$ of $G_n$ as follows:

  $ G_1^{ \otimes i} \otimes \left\{ {I,X} \right\} \otimes G_1^{n - i - 1},  $

  $ G_1^{ \otimes i} \otimes \left\{ {I,Y} \right\} \otimes G_1^{n - i - 1},  $

  $ G_1^{ \otimes i} \otimes \left\{ {I,Z} \right\} \otimes G_1^{n - i - 1},  $
\\
where $0 \le i \le n - 1$. For example, it can easily obtain
the following 32-order subgroups of $G_3$:

  $ G_3^1\left( {32} \right) = {G_2} \otimes \left\{ {I,X} \right\},  $

  $ G_3^2\left( {32} \right) = {G_2} \otimes \left\{ {I,Y} \right\},  $

  $ G_3^3\left( {32} \right) = {G_2} \otimes \left\{ {I,Z} \right\},  $

  $ G_3^4\left( {32} \right) = \left\{ {I,X} \right\} \otimes {G_2},  $

  $ G_3^5\left( {32} \right) = \left\{ {I,Y} \right\} \otimes {G_2},  $

  $ G_3^6\left( {32} \right) = \left\{ {I,Z} \right\} \otimes {G_2},  $

  $ G_3^7\left( {32} \right) = {G_1} \otimes \left\{ {I,X} \right\} \otimes {G_1},  $

  $ G_3^8\left( {32} \right) = {G_1} \otimes \left\{ {I,Y} \right\} \otimes {G_1},  $

  $ G_3^9\left( {32} \right) = {G_1} \otimes \left\{ {I,Z} \right\} \otimes {G_1}.   $
\\
However, the authors declared the above sets of subgroups are not complete. For example, the
following 8-order subgroups of $G_2$, are not in above sets,

$\begin{array}{l}G_2^7\left( 8 \right) = \left\{ I \otimes I,I \otimes Z,Z \otimes I,Z \otimes Z,X \otimes X,Y \otimes X,X \otimes Y,Y \otimes Y \right\}
\end{array}$

$\begin{array}{l}G_2^8\left( 8 \right) = \left\{ I \otimes I,Z \otimes Z,X \otimes Y,Y \otimes X,I \otimes X,
 Z \otimes Y,Y \otimes I,X \otimes Z \right\}\end{array}$

$\begin{array}{l}G_2^9\left( 8 \right) = \left\{ I \otimes I,Z \otimes Z,X \otimes Y,Y \otimes X,X \otimes I,
 Y \otimes Z,Z \otimes X,I \otimes Y \right\}\end{array}$

$\begin{array}{l}G_2^{10}\left( 8 \right) = \left\{ I \otimes I,X \otimes I,I \otimes X,X \otimes X,Z \otimes Z,
 Y \otimes Z,Z \otimes Y,Y \otimes Y \right\}\end{array}$

$\begin{array}{l}G_2^{11}\left( 8 \right) = \left\{ I \otimes I,Y \otimes I,I \otimes Y,Y \otimes Y,Z \otimes Z,
 Z \otimes X,X \otimes Z,X \otimes X \right\}.\end{array}$

Next, all the unitary operators of a subgroup are applied on the given quantum state
and the orthogonality of the output states are checked in quantum dialogue. If the
output states are mutually orthogonal, then the corresponding subgroup of
unitary operators can be used to implement quantum dialogue using
the given quantum state. Based on the mutually-orthogonal-state distinguishable principle,
 they list specific quantum states and corresponding operator sets that may be used to
quantum dialogue which can be seen in Tab.~\ref{tab8}.

\begin{table}[htbp]
\caption{List of some quantum states and corresponding appropriate operator sets}
\label{tab8}
\scalebox{1}[1]{
\tabcolsep 10pt 
\begin{tabular}{ll}
\hline\noalign{\smallskip}
Quantum state &Unitary operator set\\
\noalign{\smallskip}\hline\noalign{\smallskip}
  2-qubit Bell state&  ${G_1}\left( {4} \right)$ \\
  3-qubit GHZ state &  $G_2^1\left( 8 \right),G_2^2\left( 8 \right),G_2^4\left( 8 \right),G_2^5\left( 8 \right)$\\
  4-qubit cluster state&  ${G_2}\left( {16} \right)$\\
  5-qubit cluster state&  $G_3^4\left( {32} \right),G_3^5\left( {32} \right),G_3^7\left( {32} \right),G_3^8\left( {32} \right)$ \\
\noalign{\smallskip}\hline
\end{tabular}
}
\end{table}

As described above, the first 6 subgroups $G_2^1,G_2^2,G_2^3,
G_2^4,G_2^5,G_2^6$ of $G_2$ are constructed by Shukla et al.'s method, but the extra subgroups $G_2^7,G_2^8,G_2^9,G_2^{10},G_2^{11}$ are just provided as supplements in Ref.~\cite{Shukla2013On}. Besides, we also find that the eleven subgroups are also incomplete,
which results in the incompleteness of the unitary operator sets for a given quantum state. Obviously, Shukla et al.'s method cannot construct all the subgroups of $G_n$ (including $G_2$ and $G_3$). That is to say, there are no feasible steps and conditions to get all subgroups.

\section{Solution of constructing unitary operator sets for maximal dense coding with a class of symmetric states}

In order to solve the problem mentioned in the above section, we propose
a solution of constructing unitary operator sets to implement maximal dense coding with a class of $t$-qubit symmetric states.
And these states have two constraints:
\\
\textbf{Constraint 1:} \emph{Every state has an even number of superposition items.}
\\
\textbf{Constraint 2:} \emph{There is at least one set of $\left\lceil {{t \over 2}} \right\rceil $ qubits whose superposition items are orthogonal to each other.}

For transmitting ${2^t}$ message through a $t$-qubit symmetric state, we need a unitary operator set ${\rm{\{ }}{{\rm{U}}_0}{\rm{,}}{{\rm{U}}_1} \cdots ,
{{\rm{U}}_{{2^{t}} - 1}}{\rm{\} }}$ to operate on $n=\left\lceil {\frac{t}{2}} \right\rceil $ qubits of the state.
Our solution consists of two
phases: (1) constructing multiplicative MGP subgroups ${\rm{\{ }}{{\rm{U}}_0}{\rm{,}}{{\rm{U}}_1} \cdots ,{{\rm{U}}_{{2^{t}} - 1}}{\rm{\} }}$ which must be a group under multiplication, and (2)
selecting appropriate unitary operator sets to operate the state and making sure the results meet the mutually-orthogonal-state distinguishable principle. In order to reduce the searching scope, a two-step
construction method and its algorithm are proposed respectively in the first phase.
In the later phase, two conditions for quantum symmetric states are given
to select appropriate unitary operator
sets from the obtained multiplicative MGP subgroups.

\subsection{Constructing multiplicative MGP subgroups}
\subsubsection{Method of constructing multiplicative MGP subgroups}

\begin{figure*}[htbp]
  \includegraphics[width=3in]{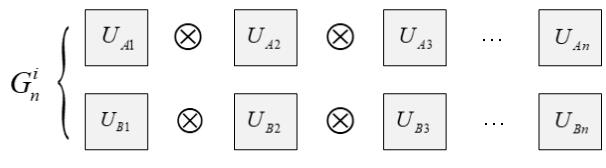}\\
  \caption{The structure of a $2^t$-order multiplicative MGP subgroup}\label{fig01}
\end{figure*}

 For the sake of more intuitive, any $2^t$-order multiplicative MGP subgroup $G_n^i$ ($i$ is the index of subgroup set)
 can be described in the form of Fig.~\ref{fig01}, where $G_n^i$ can be
 divided into two lines. Obviously, $G_n^i = \left({U_{A1}} \otimes {U_{A2}} \otimes  \cdots  \otimes {U_{An}}\right)  \cup \left({U_{B1}} \otimes {U_{B2}}
 \otimes  \cdots  \otimes {U_{Bn}}\right)$.

Through analyzing their structural characteristics of multiplicative MGP subgroups that are
transformed into the forms of Fig.~\ref{fig01}, we propose a
two-step method to construct the multiplicative MGP subgroups.

\textbf{Step 1:} \emph{Select $k=n-2$ columns from $G_n^i$ in Fig.~\ref{fig01}, and fill them with ${G_1}$,}

\begin{equation}
\label{eqn:02}
\begin{array}{l}
\left. \begin{matrix}
  {U_{Ak_1}} \otimes {U_{Ak_2}} \otimes  \cdots  \otimes {U_{Ak_{n - 2}}} \hfill \cr
  {U_{Bk_1}} \otimes {U_{Bk_2}} \otimes  \cdots  \otimes {U_{Bk_{n - 2}}} \hfill \cr
  \end{matrix}\;
\right|{U_{Ak_i}} = {U_{Bk_i}} = {G_1},
\end{array}
\end{equation}
where $\left( {i \in \left\{ {1,2,3, \cdots n - 2} \right\}} \right)$.

\begin{figure*}[htbp]
  \includegraphics[width=4.5in]{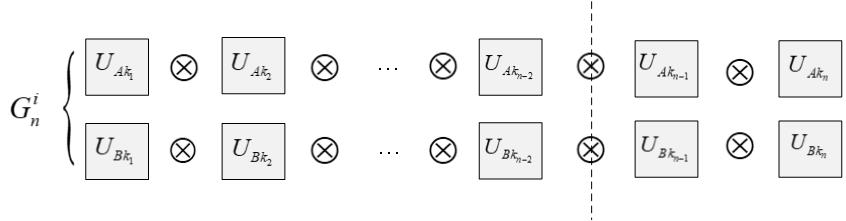}\\
  \caption{The structure division of $G_n^i$}\label{fig05}
\end{figure*}

For the sake of brevity, we put $ {U_{Ak_1}}, {U_{Ak_2}}, \cdots, {U_{Ak_{n - 2}}}$ and ${U_{Bk_1}}, {U_{Bk_2}}, \cdots, {U_{Bk_{n - 2}}} $
together, place the remainders (i.e., $ {U_{Ak_{n-1}}}, {U_{Ak_n}}$ and ${U_{Bk_{n-1}}}, {U_{Bk_n}}$)
 in the end, and get the structure of Fig.~\ref{fig05}.
Obviously, the left side of the dotted line in Fig.~\ref{fig05} are $n-2$ columns filled with ${G_1}$.

\textbf{Step 2:} \emph{The remaining 2 columns are processed as Eq.~\ref{eqn:03} and Eq.~\ref{eqn:06}}

If $t{\rm{ = 2}}n{\rm{ - 1}}$, there are two situations to construct the multiplicative MGP subgroups:
\begin{equation}
\label{eqn:03}
\begin{array}{*{20}{l}}
\left\{ \begin{array}{l}
(1)\begin{array}{l}{{U_{A{k_{n - 1}}}}{\rm{ = }}{U_{B{k_{n - 1}}}} = \left\{ {I,X} \right\},\left\{ {I,Y} \right\}{\rm{or}}\left\{ {I,Z} \right\}}\cr
{{U_{A{k_n}}}{\rm{ = }}\left\{ {I,X} \right\}\left\{ {I,Y} \right\}{\rm{or}}\left\{ {I,Z} \right\},{U_{B{k_n}}} = {G_1} - {U_{A{k_n}}}}\cr
s = n - 1
\end{array}\cr
\cr
(2)\begin{array}{l}{{U_{A{k_{n - 1}}}},{U_{A{k_n}}} = \left\{ {I,X} \right\},\left\{ {I,Y} \right\}or\left\{ {I,Z} \right\}}\cr
{{U_{B{k_{n - 1}}}} = {G_1} - {U_{A{k_{n - 1}}}},{U_{B{k_n}}} = {G_1} - {U_{A{k_n}}}}\cr
s = n - 2
\end{array}
\end{array} \right.
\cr\cr

\end{array}
\end{equation}

If $t{\rm{ = 2}}n$, there is only one situation to construct the multiplicative MGP subgroups:

\begin{equation}
\label{eqn:06}
\begin{array}{*{20}{l}}
\left\{\begin{array}{l}
{U_{A{k_{n - 1}}}}{\rm{ = }}{U_{B{k_{n - 1}}}} = {G_1}\cr
{U_{A{k_n}}}{\rm{ = }}\left\{ {I,X} \right\}\left\{ {I,Y} \right\}{\rm{or}}\left\{ {I,Z} \right\},{U_{B{k_n}}} = {G_1} - {U_{A{k_n}}}\cr
s = n - 1
\end{array}\right.
\end{array}
\end{equation}
where $s$ is the number of equal columns in $G_n^i$.

The remaining 2 columns (in the right side
of the dotted line in Fig.~\ref{fig05}) are processed according to the parity of $t$. To be specific,
when $t$ is odd (i.e., $t{\rm{ = 2}}n{\rm{ - 1}}$), there are two situations to construct the multiplicative MGP subgroups: one is
${U_{A{k_{n - 1}}}} = {U_{B{k_{n - 1}}}}$, ${U_{A{k_n}}} \cup {U_{B{k_n}}} = {G_1}$,
then we can get $s = n - 1$; the other is ${U_{A{k_{n-1}}}} \cup {U_{B{k_{n-1}}}} = {G_1}$, ${U_{A{k_n}}} \cup {U_{B{k_n}}} = {G_1}$,
and $s = n - 2$. But when $t$ is even (i.e., $t{\rm{ = 2}}n$), there is only one situation: ${U_{A{k_{n - 1}}}} =
{U_{B{k_{n - 1}}}}= {G_1}$, ${U_{A{k_n}}} \cup {U_{B{k_n}}} = {G_1}$, and obtain $s = n - 1$.

The above two steps can be more formalized into the form of Algorithm~\ref{alg1}.
After the execution of the algorithm, we can get a set $\left\{G_n^i\right\}$.
But the size of the set, i.e., the number of multiplicative MGP subgroups we obtained,
is the topic we are interested in. In fact, it can be deduced from Eq.~\ref{eqn:02} and
Eq.~\ref{eqn:03}. Define $\lambda $ as the number of $G_n^i$, if $t$ is odd (i.e., $t=2n-1$),
\begin{equation}
\label{eqn:04}
\begin{array}{*{20}{c}}
\lambda = \left\{ \begin{array}{l}
3 * C_1^1{\rm{=}}3{\rm{,}}\;\;n = 1\\
{\rm{3*}}C_{\rm{2}}^1{\rm{ + 3*3*}}C_{\rm{2}}^{\rm{2}}{\rm{=}}15{\rm{ ,}}\;\;n = 2\\
{C_n^{n - 2}\left( {3*C_2^1 + 3*3*C_2^2} \right){\rm{=}}{{n\left( {n - 1} \right)} \over 2}{\rm{*}}15,\;\;n > 2}
\end{array} \right.,
\end{array}
\end{equation}
and if $t$ is even (i.e., $t=2n$), $\lambda{\rm{ = }}1$.
So, the obtained multiplicative MGP subgroups are $\left\{ {G_n^i\left| {1 \le i \le \lambda } \right.} \right\}$.

\begin{algorithm}[h]
\renewcommand{\algorithmicrequire}{\textbf{Input:}}
\renewcommand{\algorithmicensure}{\textbf{Output:}}
\footnotesize
\caption{Constructing multiplicative MGP subgroups}
\label{alg1}
\begin{algorithmic}[1]
    \REQUIRE ${{G}_1}{\rm{ = }}\left\{ {I,X,{\rm{Y}},Z} \right\},{t}$;
    \ENSURE $\left\{G_n^i\right\}$;
    \STATE $ n = \left\lceil {\frac{t}{2}} \right\rceil$;
    \STATE $i=1$;
    \STATE $//$Select $n-2$ columns filled with ${{G}_1}$ (${U_{A{k_1}}}, {U_{A{k_2}}}, \cdots, {U_{A{k_{n - 2}}}}$ and ${U_{B{k_1}}}, {U_{B{k_2}}}, \cdots, {U_{B{k_{n - 2}}}}$);
    \FOR {$r=1$;$ $$r \leq n-2 $;$ $$r++$}
      \STATE ${U_{Ak_{r}}}{\rm{ = }}{{G}_1}$;
      \STATE ${U_{Bk_{r}}}{\rm{ = }}{{G}_1}$;
    \ENDFOR
    \STATE $//$Deal with the remaining 2 columns;
    \FOR {$x=n-1$;$x \leq n $;$x++$}
      \FOR {$y=n-1$;$y \ne x$ and $y \leq n $;$ $$y++$}
         \IF{$t$ is odd}
            \FOR {$set $ $in $ $\left\{ {\left\{ {I,X} \right\},\left\{ {I,Y} \right\},\left\{ {I,Z} \right\}} \right\}$}
               \STATE${U_{Ak_x}} \Leftarrow set$;
               \STATE${U_{Ak_y}} \Leftarrow set$;
               \STATE ${U_{Bk_x}}{\rm{ = }}{U_{Ak_x}}$;
               \STATE ${U_{Bk_y}}{\rm{ = }}{{\rm{G}}_1} - {U_{Ak_y}}$;
               \STATE Output $G_n^i={U_{Ak_1}} \otimes {U_{Ak_2}} \otimes  \cdots  \otimes {U_{Ak_n}} \cup {U_{Bk_1}} \otimes {U_{Bk_2}} \otimes  \cdots  \otimes {U_{Bk_n}}$;
               \STATE $i++$;
            \ENDFOR
            \FOR {$set $ $in$ $\left\{ {\left\{ {I,X} \right\},\left\{ {I,Y} \right\},\left\{ {I,Z} \right\}} \right\}$}
                \STATE${U_{Ak_x}} \Leftarrow set$;
                \STATE${U_{Ak_y}} \Leftarrow set$;
                \STATE ${U_{Bk_x}}{\rm{ = }}{{G}_1} - {U_{Ak_x}}$;
                \STATE ${U_{Bk_y}}{\rm{ = }}{{G}_1} - {U_{Ak_y}}$;
                \STATE Output $G_n^i= {U_{Ak_1}} \otimes {U_{Ak_2}} \otimes  \cdots  \otimes {U_{Ak_n}} \cup {U_{Bk_1}} \otimes {U_{Bk_2}} \otimes  \cdots  \otimes {U_{Bk_n}}$;
                \STATE $i++$;
            \ENDFOR
         \ELSE \STATE $//$t is even;
            \STATE ${U_{Ak_x}}{\rm{ = }}{{G}_1}$;
            \STATE ${U_{Bk_x}}{\rm{ = }}{{G}_1}$;
            \FOR {$set $ $in $ $\left\{ {\left\{ {I,X} \right\},\left\{ {I,Y} \right\},\left\{ {I,Z} \right\}} \right\}$}
              \STATE ${U_{Ak_y}} \Leftarrow set$;
              \STATE ${U_{Bk_y}}{\rm{ = }}{{G}_1} - {U_{Ak_y}}$;
              \STATE Output $G_n^i= {U_{Ak_1}} \otimes {U_{Ak_2}} \otimes  \cdots  \otimes {U_{Ak_n}} \cup {U_{Bk_1}} \otimes {U_{Bk_2}} \otimes  \cdots  \otimes {U_{Bk_n}}$;
              \STATE $i++$;

            \ENDFOR
         \ENDIF

\ENDFOR
\ENDFOR
\end{algorithmic}
\end{algorithm}

\subsubsection{Examples}
Taking $n=2$ and $n=3$ as examples, we demonstrate our construction method in detail as follows.\\

\textbf{(1)	$n=2$}

\begin{figure}[htbp]
  \includegraphics[width=1.2in]{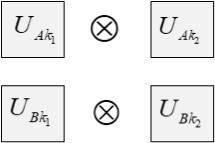}\\
  \caption{The structure of a 2-qubit multiplicative MGP subgroup}\label{fig02}
\end{figure}

When $n=2$ (i.e., we select 2 qubits to operate), so $k=n-2=0$,
which means there is no columns to be filled with $G_1$ in Step 1. Then referring to Step 2 (also Eq.~\ref{eqn:03}),
there are two possible situations: $t = 2n - 1 = 3$ (i.e., 3-qubit state),
$t = 2n = 4$ (i.e., 4-qubit state), and their multiplicative MGP subgroups can be described in the form of Fig.~\ref{fig02} as Eq.~\ref{eqn:05} and Eq.~\ref{eqn:16}.

If $t{\rm{ = }}3$, there are two situations:

\begin{equation}
\label{eqn:05}
\begin{array}{l}
\left\{ \begin{array}{l}
(1)\begin{array}{l}{U_{A{k_1}} = U_{B{k_1}} = \left\{ {I,X} \right\},\left\{ {I,Y} \right\}or\left\{ {I,Z} \right\}}\cr
U_{A{k_2}} = \left\{ {I,X} \right\}\left\{ {I,Y} \right\}or\left\{ {I,Z} \right\}, U_{B{k_2}} = {G_1} - {U_{A{k_2}}}\cr
s = 1\cr
\end{array}\\
\cr
(2)\begin{array}{l}{U_{A{k_1}},U_{A{k_2}}=\left\{ {I,X} \right\},\left\{ {I,Y} \right\}or\left\{ {I,Z} \right\}}\cr
{U_{B{k_1}} = {G_1} - {U_{A{k_1}}},U_{B{k_2}} = {G_1} - {U_{A{k_2}}}}\cr
s = 0
\end{array}
\end{array} \right.\\
\end{array}
\end{equation}

If $t{\rm{ = }}4$:
\begin{equation}
\label{eqn:16}
\begin{array}{l}
\left\{\begin{array}{l}
{U_{A{k_1}}}{\rm{ = }}{U_{{B{k_1}}}}{\rm{ = }}{G_1}\cr
U_{A{k_2}} = \left\{ {I,X} \right\}\left\{ {I,Y} \right\}or\left\{ {I,Z} \right\}, U_{B{k_2}} = {G_1} - {U_{A{k_2}}}\cr
s = 1\\
\end{array}\right.
\end{array}
\end{equation}
\\here, ${k_1} \ne {k_2}$ and $k_1,k_2 \in \left\{ {1,2} \right\}$.

For 3-qubit state (i.e., $t=3$), we need an 8-order multiplicative MGP subgroups in quantum maximal dense coding.
It can be decomposed into the union of two 4-order MGP subgroups in the  form of Fig.~\ref{fig02}.
According to Eq.~\ref{eqn:04}, there are 15 multiplicative MGP subgroups exist for 3-qubit state:

$G_2^1\left( 8 \right){\rm{ = }}\left( {\left\{ {I,\left. X \right\} \otimes \left\{ {\left. {I,X} \right\}} \right.} \right.} \right) \cup \left( {\left\{ {Y,\left. Z \right\} \otimes \left\{ {\left. {I,X} \right\}} \right.} \right.} \right)$

$G_2^2\left( 8 \right){\rm{ = }}\left( {\left\{ {I,\left. X \right\} \otimes \left\{ {\left. {I,Y} \right\}} \right.} \right.} \right) \cup \left( {\left\{ {Y,\left. Z \right\} \otimes \left\{ {\left. {I,Y} \right\}} \right.} \right.} \right)$

$G_2^3\left( 8 \right){\rm{ = }}\left( {\left\{ {I,\left. X \right\} \otimes \left\{ {\left. {I,Z} \right\}} \right.} \right.} \right) \cup \left( {\left\{ {Y,\left. Z \right\} \otimes \left\{ {\left. {I,Z} \right\}} \right.} \right.} \right)$

$G_2^4\left( 8 \right){\rm{ = }}\left( {\left\{ {I,\left. X \right\} \otimes \left\{ {\left. {I,Y} \right\}} \right.} \right.} \right) \cup \left( {\left\{ {I,\left. X \right\} \otimes \left\{ {\left. {X,Z} \right\}} \right.} \right.} \right)$

$G_2^5\left( 8 \right){\rm{ = }}\left( {\left\{ {I,\left. {Y} \right\} \otimes \left\{ {\left. {I,Y} \right\}} \right.} \right.} \right) \cup \left( {\left\{ {I,\left. {Y} \right\} \otimes \left\{ {\left. {X,Z} \right\}} \right.} \right.} \right)$

$G_2^6\left( 8 \right){\rm{ = }}\left( {\left\{ {I,\left. Z \right\} \otimes \left\{ {\left. {I,Y} \right\}} \right.} \right.} \right) \cup \left( {\left\{ {I,\left. Z \right\} \otimes \left\{ {\left. {X,Z} \right\}} \right.} \right.} \right)$

$G_2^7\left( 8 \right){\rm{ = }}\left( {\left\{ {I,\left. Z \right\} \otimes \left\{ {\left. {I,Z} \right\}} \right.} \right.} \right) \cup \left( {\left\{ {X,\left. {Y} \right\} \otimes \left\{ {\left. {X,Y} \right\}} \right.} \right.} \right)$

$G_2^8\left( 8 \right){\rm{ = }}\left( {\left\{ {I,\left. {Y} \right\} \otimes \left\{ {\left. {I,X} \right\}} \right.} \right.} \right) \cup \left( {\left\{ {X,\left. Z \right\} \otimes \left\{ {\left. {Z,Y} \right\}} \right.} \right.} \right)$

$G_2^9\left( 8 \right){\rm{ = }}\left( {\left\{ {I,\left. X \right\} \otimes \left\{ {\left. {I,Y} \right\}} \right.} \right.} \right) \cup \left( {\left\{ {Z,\left. {Y} \right\} \otimes \left\{ {\left. {Z,X} \right\}} \right.} \right.} \right)$

$G_2^{10}\left( 8 \right){\rm{ = }}\left( {\left\{ {I,\left. X \right\} \otimes \left\{ {\left. {I,X} \right\}} \right.} \right.} \right) \cup \left( {\left\{ {Z,\left. {Y} \right\} \otimes \left\{ {\left. {Z,Y} \right\}} \right.} \right.} \right)$

$G_2^{11}\left( 8 \right){\rm{ = }}\left( {\left\{ {I,\left. {Y} \right\} \otimes \left\{ {\left. {I,Y} \right\}} \right.} \right.} \right) \cup \left( {\left\{ {Z,\left. X \right\} \otimes \left\{ {\left. {Z,X} \right\}} \right.} \right.} \right)$

$G_2^{12}\left( 8 \right){\rm{ = }}\left( {\left\{ {I,\left. X \right\} \otimes \left\{ {\left. {I,Z} \right\}} \right.} \right.} \right) \cup \left( {\left\{ {Y,\left. Z \right\} \otimes \left\{ {\left. {Y,X} \right\}} \right.} \right.} \right)$

$G_2^{13}\left( 8 \right){\rm{ = }}\left( {\left\{ {I,\left. Z \right\} \otimes \left\{ {\left. {I,X} \right\}} \right.} \right.} \right) \cup \left( {\left\{ {Y,\left. X \right\} \otimes \left\{ {\left. {Y,Z} \right\}} \right.} \right.} \right)$

$G_2^{14}\left( 8 \right){\rm{ = }}\left( {\left\{ {I,\left. {Y} \right\} \otimes \left\{ {\left. {I,Z} \right\}} \right.} \right.} \right) \cup \left( {\left\{ {X,\left. Z \right\} \otimes \left\{ {\left. {Y,X} \right\}} \right.} \right.} \right)$

$G_2^{15}\left( 8 \right){\rm{ = }}\left( {\left\{ {I,\left. Z \right\} \otimes \left\{ {\left. {I,Y} \right\}} \right.} \right.} \right) \cup \left( {\left\{ {X,\left. {Y} \right\} \otimes \left\{ {\left. {Z,X} \right\}} \right.} \right.} \right)$
\\Here, $G_2^1,G_2^2,G_2^3,G_2^4,G_2^5,G_2^6$ belong to the case of $s = 1$, and $G_2^7,G_2^8,G_2^9,G_2^{10},G_2^{11},G_2^{12},G_2^{13},
  G_2^{14},G_2^{15}$ belong to the case of $s = 0$.

For 4-qubit state (i.e., $t=4$), we need a 16-order
 multiplicative MGP subgroups in quantum maximal dense coding. It can be decomposed into the union of
two 8-order MGP subgroups in the form of Fig.~\ref{eqn:02} as follows:

$\begin{array}{l}
G_2^1\left( {16} \right) = \left\{ {\left\{ {I,X,Y,Z} \right\} \otimes \left\{ {I,X} \right\}} \right\}\cup \left\{ {\left\{ {I,X,Y,Z} \right\} \otimes \left\{ {Y,Z} \right\}} \right\}\end{array}$

$\begin{array}{l}
G_2^2\left( {16} \right) = \left\{ {\left\{ {I,X,Y,Z} \right\} \otimes \left\{ {I,Y} \right\}} \right\}\cup \left\{ {\left\{ {I,X,Y,Z} \right\} \otimes \left\{ {X,Z} \right\}} \right\}\end{array}$

$\begin{array}{l}
G_2^3\left( {16} \right) = \left\{ {\left\{ {I,X,Y,Z} \right\} \otimes \left\{ {I,Z} \right\}} \right\}\cup \left\{ {\left\{ {I,X,Y,Z} \right\} \otimes \left\{ {X,Y} \right\}} \right\}\end{array}$.
\\Actually, the above three subgroups are the same:
${G_2}\left( {16} \right)=\left\{ {\left\{I,X,Y,Z\right\} \otimes \left\{I,X,Y,Z\right\}} \right\}$.

\textbf{(2) $n=3$}

\begin{figure}[htbp]
  \includegraphics[width=2in]{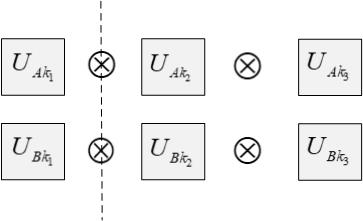}\\
  \caption{The structure of a 3-qubit multiplicative MGP subgroup}\label{fig03}
\end{figure}

When $n=3$, so $k=n-2=1$,
which means $U_{Ak_1} = U_{Bk_1} = G_1$ ($k_1 \in \left\{ {1,2,3} \right\}$). Then referring to Step 2,
there are two possible situations: $t = 2n - 1 = 5$ (i.e., 5-qubit state),
$t = 2n = 6$ (i.e., 6-qubit state), and their specific multiplicative MGP subgroups
can be described in the form of Fig.~\ref{fig03} as Eq.~\ref{eqn:15} and Eq.~\ref{eqn:17}.

If $t{\rm{ = }}5$, there are two situations:
\begin{equation}
\label{eqn:15}
\begin{array}{l}
\left\{ \begin{array}{l}
(1)\begin{array}{l}{U_{Ak_2} = U_{Bk_2} = \left\{ {I,X} \right\},\left\{ {I,Y} \right\}or\left\{ {I,Z} \right\}}\cr
{U_{Ak_3}=\left\{ {I,X} \right\}\left\{ {I,Y} \right\}or\left\{ {I,Z} \right\}, U_{Bk_3} = {G_1}- U_{Ak_3}}\cr
s = 2
\end{array}\\
\cr
(2)\begin{array}{l}{U_{Ak_2},U_{Ak_3}=\left\{ {I,X} \right\},\left\{ {I,Y} \right\}or\left\{ {I,Z} \right\}}\cr
{U_{Bk_2} = {G_1}-U_{Ak_2},U_{Bk_3} = {G_1}-U_{Ak_3}}\cr
here,\;s=1
\end{array}
\end{array} \right.
\end{array}
\end{equation}

If $t{\rm{ = }}6$,
\begin{equation}
\label{eqn:17}
\begin{array}{l}
\left\{\begin{array}{l}
{U_{Ak_2}}{\rm{ = }}{U_{{Bk_2}}}{\rm{ = }}{G_1}\cr
U_{Ak_3}=\left\{ {I,X} \right\}\left\{ {I,Y} \right\}or\left\{ {I,Z} \right\}, U_{Bk_3} ={G_1}-U_{Ak_3}\cr
here,\;s = 2
\end{array}\right.
\end{array}
\end{equation}
\\here, ${k_1} \ne {k_2} \ne {k_3}$ and $k_2,k_3 \in \left\{ {1,2,3} \right\}$.

For 5-qubit state (i.e., $t=5$), a 32-order multiplicative MGP subgroups is required in maximal dense coding.
It can be decomposed into the union of two 16-order MGP subgroups in the form of Fig.~\ref{fig03}.
According to Eq.~\ref{eqn:04}, there are 45 multiplicative MGP subgroups exist for 3-qubit unitary operator sets.
Here, because the number of 45 is too large, we only list 15 multiplicative MGP subgroups in
which $G_1$ is in the last column. Then, we can quickly list them as follows:

$G_3^1\left( {32} \right){\rm{ = }}G_2^1 \otimes {G_1}$

$G_3^2\left( {32} \right){\rm{ = }}G_2^2 \otimes {G_1}$

$G_3^3\left( {32} \right){\rm{ = }}G_2^3 \otimes {G_1}$

$G_3^4\left( {32} \right){\rm{ = }}G_2^4 \otimes {G_1}$

$G_3^5\left( {32} \right){\rm{ = }}G_2^5 \otimes {G_1}$

$G_3^6\left( {32} \right){\rm{ = }}G_2^6 \otimes {G_1}$

$G_3^7\left( {32} \right){\rm{ = }}G_2^7 \otimes {G_1}$

$G_3^8\left( {32} \right){\rm{ = }}G_2^8 \otimes {G_1}$

$G_3^9\left( {32} \right){\rm{ = }}G_2^9 \otimes {G_1}$

$G_3^{10}\left( {32} \right){\rm{ = }}G_2^{10} \otimes {G_1}$

$G_3^{11}\left( {32} \right){\rm{ = }}G_2^{11} \otimes {G_1}$

$G_3^{12}\left( {32} \right){\rm{ = }}G_2^{12} \otimes {G_1}$

$G_3^{13}\left( {32} \right){\rm{ = }}G_2^{13} \otimes {G_1}$

$G_3^{14}\left( {32} \right){\rm{ = }}G_2^{14} \otimes {G_1}$

$G_3^{15}\left( {32} \right){\rm{ = }}G_2^{15} \otimes {G_1}$
\\Here, $G_3^1,G_3^2,G_3^3,G_3^4,G_3^5,G_3^6$ belong to the case of $s = 2$, and $G_3^7,G_3^8,G_3^9,G_3^{10},
G_3^{11},G_3^{12},G_3^{13},G_3^{14},G_3^{15}$ belong to the case of $s = 1$.

For 6-qubit state (i.e., $t=6$), we need a 64-order
 multiplicative MGP subgroups in quantum maximal dense coding. It can be decomposed into the union of
two 32-order MGP subgroups in the form of Fig.~\ref{fig03} as follows:

$\begin{array}{l}
G_3^1\left( {64} \right) = \left\{ {\left\{ {I,X,Y,Z} \right\} \otimes\left\{ {I,X,Y,Z} \right\} \otimes \left\{ {I,X} \right\}} \right\}\cup \left\{ {\left\{ {I,X,Y,Z} \right\} \otimes\left\{ {I,X,Y,Z} \right\} \otimes \left\{ {Y,Z} \right\}} \right\}\end{array}$

$\begin{array}{l}
G_3^2\left( {64} \right) = \left\{ {\left\{ {I,X,Y,Z} \right\} \otimes\left\{ {I,X,Y,Z} \right\} \otimes \left\{ {I,Y} \right\}} \right\}\cup \left\{ {\left\{ {I,X,Y,Z} \right\} \otimes\left\{ {I,X,Y,Z} \right\} \otimes \left\{ {X,Z} \right\}} \right\}\end{array}$

$\begin{array}{l}
G_3^3\left( {64} \right) = \left\{ {\left\{ {I,X,Y,Z} \right\} \otimes\left\{ {I,X,Y,Z} \right\} \otimes \left\{ {I,Z} \right\}} \right\}\cup \left\{ {\left\{ {I,X,Y,Z} \right\} \otimes\left\{ {I,X,Y,Z} \right\} \otimes \left\{ {X,Y} \right\}} \right\}
\end{array}$.
\\Obviously, they can be expressed as below:\\
${G_3}\left( {64} \right)=\left\{ {\left\{{I,X,Y,Z}\right\} \otimes \left\{{I,X,Y,Z}\right\} \otimes \left\{ {I,X,Y,Z} \right\}} \right\}$.
\subsection{Selecting appropriate unitary operator sets}
\subsubsection{Two conditions to select the appropriate unitary operator sets}

In the previous section, we demonstrate how to construct the multiplicative MGP subgroups. Compared to the original exhaustive method, this method can greatly reduce the scale of calculation. Next, we need to select appropriate unitary operator sets from the above MGP subgroups. For a specific quantum state, there are two conditions as follows.
\\
\textbf{Condition 1:} \emph{In each MGP subgroup $G_n^{i}$,
the elements with even number of $Z$ are not allowed.}

Because the elements
with even number of $Z$ have the same operational effect as the unit cell in the subgroup (i.e., ${I^{ \otimes n}}$). So
if any element of $G_n^{i}$ has even number of $Z$, $G_n^{i}$ is unable to achieve maximal dense coding.
For 3-qubit quantum states, we need 2-qubit unitary operators and the element $I \otimes I$ and $Z \otimes Z$ have the
same effect, which help us exclude $G_2^3,G_2^6,G_2^7,G_2^8,G_2^9,G_2^{10},G_2^{11}$.
\\
\textbf{Condition 2:} \emph{For a specific quantum state, we select $\left\lceil {{t \over 2}} \right\rceil $ qubits to operate and the product states of operated qubits must be mutually orthogonal.}

The product states of operated qubits can consist of a multi-qubit state set. In order to meet the requirement of Condition 2, all the
product states in the set must be mutually orthogonal.
Taking one of 4-qubit cluster states, $\left| {{c}} \right\rangle_4  = {1 \over 2}{\left( {\left| {0000} \right\rangle  + \left| {0011}
\right\rangle  + \left| {1100} \right\rangle  - \left| {1111} \right\rangle } \right)_{1234}}$, as an example, if we choose the
1st and 2nd qubits as operated qubits, the product states of these two qubits consist of the set
$\left\{ {\left| {00} \right\rangle, \left| {00} \right\rangle,
\left| {11} \right\rangle, \left| {11} \right\rangle } \right\}$. Obviously, the elements of the set are not
mutually orthogonal, so this encoding method cannot achieve maximal dense coding.
However, if we operate the 1st and 3rd qubits, then the 2-qubit product state set is $\left\{ {\left| {00} \right\rangle, \left| {01} \right\rangle,
\left| {10} \right\rangle, \left| {11} \right\rangle } \right\}$, so the encoding method can work because
 all the elements are
mutually orthogonal.  Therefore,
the positions of operated qubit for $\left| {{c}} \right\rangle_4$ are relatively deterministic, i.e.,
one of the operated qubits must
be chosen from qubit 1 or 2, and the other is qubit 3 or 4.

The above two conditions enable us to select the appropriate unitary operator sets and feasible encoding method with regard to a certain
quantum state. To verify its correctness, we take 2-qubit Bell, 3-qubit GHZ, 4-qubit cluster and 5-qubit cluster states as examples,
and demonstrate how to select all appropriate unitary operator sets.
\subsubsection{Examples}
Taking 3-qubit GHZ state, 4-qubit W, 4-qubit cluster and 5-qubit cluster states as quantum resource in maximal dense coding,
we demonstrate that how to find all appropriate unitary operator sets through the above solution.

\textbf{(1) 3-qubit GHZ state }

In the area of quantum information theory, a GHZ (Greenberger-Horne-Zeilinger~\cite{Greenberger1989Going}) state is an entangled quantum state of $M > 2$ subsystems. In the case of each subsystem being two-dimensional,
that is for qubits, it reads
\begin{equation}
\label{eqn:08}
\left| {GHZ} \right\rangle  = {{{{\left| 0 \right\rangle }^{ \otimes M}} + {{\left| 1 \right\rangle }^{ \otimes M}}} \over {\sqrt 2 }}.
\end{equation}
The simplest one is the 3-qubit GHZ state:
\begin{equation}
\label{eqn:14}
\left| {GHZ} \right\rangle_3 {\rm{ = }}{{\left( {\left| {000} \right\rangle {\rm{ + }}\left| {111} \right\rangle } \right)} \over {\sqrt 2 }}.
\end{equation}

Taking $\left| {GHZ} \right\rangle_3 $ as an example, after the construction phase,
the multiplicative MGP subgroups are shown as follows,

$\begin{array}{l}G_2^1\left( 8 \right){\rm{ = }}\left( {\left\{ {I,\left. X \right\} \otimes \left\{ {\left. {I,X} \right\}} \right.} \right.} \right) \cup \left( {\left\{ {Y,\left. Z \right\} \otimes \left\{ {\left. {I,X} \right\}} \right.} \right.} \right)\\
\;\;\;\;\;\;\;\;\;\; =\left\{ {I \otimes I,I \otimes X,X \otimes I,X \otimes X, Y \otimes I,Y \otimes X,Z \otimes I,Z \otimes X} \right\}
\end{array}$

$\begin{array}{l}G_2^2\left( 8 \right){\rm{ = }}\left( {\left\{ {I,\left. X \right\} \otimes \left\{ {\left. {I,Y} \right\}} \right.} \right.} \right) \cup \left( {\left\{ {Y,\left. Z \right\} \otimes \left\{ {\left. {I,Y} \right\}} \right.} \right.} \right)\\
\;\;\;\;\;\;\;\;\;\;=\left\{ {I \otimes I,I \otimes Y,X \otimes I,X \otimes Y,Y \otimes I,Y \otimes Y,Z \otimes I,Z \otimes Y} \right\}
\end{array}$

$\begin{array}{l}G_2^3\left( 8 \right){\rm{ = }}\left( {\left\{ {I,\left. X \right\} \otimes \left\{ {\left. {I,Z} \right\}} \right.} \right.} \right) \cup \left( {\left\{ {Y,\left. Z \right\} \otimes \left\{ {\left. {I,Z} \right\}} \right.} \right.} \right)\\
\;\;\;\;\;\;\;\;\;\;=\left\{ {I \otimes I,I \otimes Z,X \otimes I,X \otimes Z,Y \otimes I,Y \otimes Z,Z \otimes I,Z \otimes Z} \right\}
\end{array}$

$\begin{array}{l}G_2^4\left( 8 \right){\rm{ = }}\left( {\left\{ {I,\left. X \right\} \otimes \left\{ {\left. {I,Y} \right\}} \right.} \right.} \right) \cup \left( {\left\{ {I,\left. X \right\} \otimes \left\{ {\left. {X,Z} \right\}} \right.} \right.} \right)\\
\;\;\;\;\;\;\;\;\;\;=\left\{ {I \otimes I,I \otimes Y,X \otimes I,X \otimes Y,I \otimes X,I \otimes Z,X \otimes X,X \otimes Z} \right\}
\end{array}$

$\begin{array}{l}G_2^5\left( 8 \right){\rm{ = }}\left( {\left\{ {I,\left. {Y} \right\} \otimes \left\{ {\left. {I,Y} \right\}} \right.} \right.} \right) \cup \left( {\left\{ {I,\left. {Y} \right\} \otimes \left\{ {\left. {X,Z} \right\}} \right.} \right.} \right)\\
\;\;\;\;\;\;\;\;\;\;=\left\{ {I \otimes I,I \otimes Y,Y \otimes I,Y \otimes Y,I \otimes X,I \otimes Z,Y \otimes X,Y \otimes Z} \right\}

\end{array}$

$\begin{array}{l}G_2^6\left( 8 \right){\rm{ = }}\left( {\left\{ {I,\left. Z \right\} \otimes \left\{ {\left. {I,Y} \right\}} \right.} \right.} \right) \cup \left( {\left\{ {I,\left. Z \right\} \otimes \left\{ {\left. {X,Z} \right\}} \right.} \right.} \right)\\
\;\;\;\;\;\;\;\;\;\;=\left\{ {I \otimes I,I \otimes Y,Z \otimes I,Z \otimes Y,I \otimes X,I \otimes Z,Z \otimes X,Z \otimes Z} \right\}

\end{array}$

$\begin{array}{l}G_2^7\left( 8 \right){\rm{ = }}\left( {\left\{ {I,\left. Z \right\} \otimes \left\{ {\left. {I,Z} \right\}} \right.} \right.} \right) \cup \left( {\left\{ {X,\left. {Y} \right\} \otimes \left\{ {\left. {X,Y} \right\}} \right.} \right.} \right)\\
\;\;\;\;\;\;\;\;\;\;=\left\{ {I \otimes I,I \otimes Z,Z \otimes I,Z \otimes Z,X \otimes X,X\otimes Y,Y \otimes X,Y \otimes Y} \right\}
\end{array}$

$\begin{array}{l}G_2^8\left( 8 \right){\rm{ = }}\left( {\left\{ {I,\left. {Y} \right\} \otimes \left\{ {\left. {I,X} \right\}} \right.} \right.} \right) \cup \left( {\left\{ {X,\left. Z \right\} \otimes \left\{ {\left. {Z,Y} \right\}} \right.} \right.} \right)\\
\;\;\;\;\;\;\;\;\;\;=\left\{ {I \otimes I,I \otimes X,Y \otimes I,Y \otimes X,X \otimes Z,X \otimes Y,Z \otimes Z,Z \otimes Y} \right\}
\end{array}$

$\begin{array}{l}G_2^9\left( 8 \right){\rm{ = }}\left( {\left\{ {I,\left. X \right\} \otimes \left\{ {\left. {I,Y} \right\}} \right.} \right.} \right) \cup \left( {\left\{ {Z,\left. {Y} \right\} \otimes \left\{ {\left. {Z,X} \right\}} \right.} \right.} \right)\\
\;\;\;\;\;\;\;\;\;\;=\left\{ {I \otimes I,I \otimes Y,X \otimes I,X \otimes Y,Z \otimes Z,Z \otimes X,Y \otimes Z,Y \otimes X} \right\}
\end{array}$

$\begin{array}{l}G_2^{10}\left( 8 \right){\rm{ = }}\left( {\left\{ {I,\left. X \right\} \otimes \left\{ {\left. {I,X} \right\}} \right.} \right.} \right) \cup \left( {\left\{ {Z,\left. {Y} \right\} \otimes \left\{ {\left. {Z,Y} \right\}} \right.} \right.} \right)\\
\;\;\;\;\;\;\;\;\;\;\;\;=\left\{ {I \otimes I,I \otimes X,X \otimes I,X \otimes X,Z \otimes Z,Z \otimes Y,Y \otimes Z,Y \otimes Y} \right\}
\end{array}$

$\begin{array}{l}G_2^{11}\left( 8 \right){\rm{ = }}\left( {\left\{ {I,\left. {Y} \right\} \otimes \left\{ {\left. {I,Y} \right\}} \right.} \right.} \right) \cup \left( {\left\{ {Z,\left. X \right\} \otimes \left\{ {\left. {Z,X} \right\}} \right.} \right.} \right)\\
\;\;\;\;\;\;\;\;\;\;\;\;=\left\{ {I \otimes I,I \otimes Y,Y \otimes I,Y \otimes Y,Z \otimes Z,Z \otimes X,X \otimes Z,X \otimes X} \right\}
\end{array}$

$\begin{array}{l}G_2^{12}\left( 8 \right){\rm{ = }}\left( {\left\{ {I,\left. X \right\} \otimes \left\{ {\left. {I,Z} \right\}} \right.} \right.} \right) \cup \left( {\left\{ {Y,\left. Z \right\} \otimes \left\{ {\left. {Y,X} \right\}} \right.} \right.} \right)\\
\;\;\;\;\;\;\;\;\;\;\;\;=\left\{ {I \otimes I,I \otimes Z,X \otimes I,X \otimes Z,Y \otimes X,Y \otimes Y,Z \otimes X,Z \otimes Y} \right\}
\end{array}$

$\begin{array}{l}G_2^{13}\left( 8 \right){\rm{ = }}\left( {\left\{ {I,\left. Z \right\} \otimes \left\{ {\left. {I,X} \right\}} \right.} \right.} \right) \cup \left( {\left\{ {Y,\left. X \right\} \otimes \left\{ {\left. {Y,Z} \right\}} \right.} \right.} \right)\\
\;\;\;\;\;\;\;\;\;\;\;\;=\left\{ {I \otimes I,I \otimes X,Z \otimes I,Z \otimes X,Y \otimes Y,Y \otimes Z,X \otimes Y,X \otimes Z} \right\}
\end{array}$

$\begin{array}{l}
G_2^{14}\left( 8 \right){\rm{ = }}\left( {\left\{ {I,\left. {Y} \right\} \otimes \left\{ {\left. {I,Z} \right\}} \right.} \right.} \right) \cup \left( {\left\{ {X,\left. Z \right\} \otimes \left\{ {\left. {Y,X} \right\}} \right.} \right.} \right)\\
\;\;\;\;\;\;\;\;\;\;\;\;=\left\{ {I \otimes I,I \otimes Z,{Y} \otimes I,{Y} \otimes Z,Z \otimes X,Z \otimes Y,X \otimes X,X \otimes Y} \right\}
\end{array}$

$\begin{array}{l}
G_2^{15}\left( 8 \right){\rm{ = }}\left( {\left\{ {I,\left. Z \right\} \otimes \left\{ {\left. {I,Y} \right\}} \right.} \right.} \right) \cup \left( {\left\{ {X,\left. {Y} \right\} \otimes \left\{ {\left. {Z,X} \right\}} \right.} \right.} \right)\\
\;\;\;\;\;\;\;\;\;\;\;\;=\left\{ {I \otimes I,I \otimes {Y},{Z} \otimes I,{Z} \otimes {Y},{X} \otimes {Z,X} \otimes {X},{Y} \otimes {Z},{Y} \otimes {X}} \right\}.
\end{array}$\\
Now, we exclude $G_2^3,G_2^6,G_2^7,G_2^8,G_2^9,G_2^{10},G_2^{11}$ who violate Condition 1, that is,
the appropriate unitary operator sets
are $G_2^1,G_2^2,G_2^4,G_2^5,G_2^{12},G_2^{13},G_2^{14},G_2^{15}$.
Further, as an example, in Tab.~\ref{tab5} we have provided group multiplication table for $G_2^{12}\left( 8 \right)$. Similar group multiplication tables and tables for maximal dense coding can easily be constructed for all other cases mentioned here. Since this verification is an easy task, we have not provided such tables here. Besides, the positions of the two operated qubits which satisfy Condition 2 are arbitrary. Then, Tab.~\ref{tab7} illustrates
that $G_2^{12}$ can be operated on qubit 1 and 2 of 3-qubit GHZ state.

\newcommand{\tabincell}[2]{
\begin{tabular}{@{}#1@{}}#2\end{tabular}
}

\begin{table*}[htbp]
\caption{Group multiplication table for $G_2^{12}$}
\label{tab5}
\scalebox{1}[1]{
\tabcolsep 8pt 
\begin{tabular}{lllllllll}
\hline\noalign{\smallskip}
Unitary operators&$I \otimes I$ & $I \otimes Z$ & $X \otimes I$&$X \otimes Z$&$Y \otimes X$&$Y \otimes Y$&$Z \otimes X$&$Z \otimes Y$\\
\noalign{\smallskip}\hline\noalign{\smallskip}
 $I \otimes I$ & $I \otimes I$ & $I \otimes Z$ & $X \otimes I$ & $X \otimes Z$ & $Y \otimes X$ & $Y \otimes Y$ & $Z \otimes X$ & $Z \otimes Y$\\
 $I \otimes Z$ & $I \otimes Z$ & $I \otimes I$ & $X \otimes Z$ & $X \otimes I$ & $Y \otimes Y$ & $Y \otimes X$ & $Z \otimes Y$ & $Z \otimes X$\\
 $X \otimes I$ & $X \otimes I$ & $X \otimes Z$ & $I \otimes I$ & $I \otimes Z$ & $Z \otimes X$ & $Z \otimes Y$ & $Y \otimes X$ & $Y \otimes Y$\\
 $X \otimes Z$ & $X \otimes Z$ & $X \otimes I$ & $I \otimes Z$ & $I \otimes I$ & $Z \otimes Y$ & $Z \otimes X$ & $Y \otimes Y$ & $Y \otimes X$\\
 $Y \otimes X$ & $Y \otimes X$ & $Y \otimes Y$ & $Z \otimes X$ & $Z \otimes Y$ & $I \otimes I$ & $I \otimes Z$ & $X \otimes I$ & $X \otimes Z$\\
 $Y \otimes Y$ & $Y \otimes Y$ & $Y \otimes X$ & $Z \otimes Y$ & $Z \otimes X$ & $I \otimes Z$ & $I \otimes I$ & $X \otimes Z$ & $X \otimes I$\\
 $Z \otimes X$ & $Z \otimes X$ & $Z \otimes Y$ & $Y \otimes X$ & $Y \otimes Y$ & $X \otimes I$ & $X \otimes Z$ & $I \otimes I$ & $I \otimes Z$\\
 $Z \otimes Y$ & $Z \otimes Y$ & $Z \otimes X$ & $Y \otimes Y$ & $Y \otimes X$ & $X \otimes Z$ & $X \otimes I$ & $I \otimes Z$ & $I \otimes I$\\
\noalign{\smallskip}\hline
\end{tabular}
}
\end{table*}

\begin{table}[htbp]
\caption{Maximal dense coding of 3-qubit GHZ state using the elements of $G_2^{12}\left( 8 \right)$}
\label{tab7}

\tabcolsep 22pt 
\begin{tabular}{ll}
\hline\noalign{\smallskip}
\tabincell{c}{ Unitary operators \\on qubits 1 and 2} &  3-qubit GHZ state \\
\noalign{\smallskip}\hline\noalign{\smallskip}
  ${U_0}$=I$ \otimes $I & $\frac{1}{2}{\rm{(\left|{000}\right\rangle{\rm{ + }}\left| {111} \right\rangle)}}$\\

  ${U_1}$=I$ \otimes $Z & $\frac{1}{2}{\rm{(\left| {000} \right\rangle{\rm{ - }}\left| {111} \right\rangle)}}$\\

  ${U_2}$=X$ \otimes $I & $\frac{1}{2}{\rm{(\left| {100} \right\rangle{\rm{ + }}\left| {011} \right\rangle)}}$\\

  ${U_3}$=X$ \otimes $Z & $\frac{1}{2}{\rm{(\left| {100} \right\rangle{\rm{ - }}\left| {011} \right\rangle)}}$\\

  ${U_4}$=Y$ \otimes $X & $\frac{1}{2}{\rm{(-\left| {110} \right\rangle{\rm{ + }}\left| {001} \right\rangle)}}$\\

  ${U_5}$=Y$ \otimes $Y & $\frac{1}{2}{\rm{(\left| {110} \right\rangle{\rm{ + }}\left| {001} \right\rangle)}}$\\

  ${U_6}$=Z$ \otimes $X & $\frac{1}{2}{\rm{(\left| {010} \right\rangle{\rm{ - }}\left| {101} \right\rangle)}}$\\

  ${U_7}$=Z$ \otimes $Y & $\frac{1}{2}{\rm{(-\left| {010} \right\rangle{\rm{ - }}\left| {101} \right\rangle)}}$\\
\noalign{\smallskip}\hline
\end{tabular}
\end{table}

\textbf{(2) 4-qubit W state}

The $W$ state~\cite{D2000Three} is one of the two non-biseparable classes of three-qubit states
(the other being the GHZ state), which can not be transformed (not even probabilistically) into each other by local quantum
operations. The notion of $W$ state has been generalized for $t$ qubits,
\begin{equation}
\label{eqn:07}
\begin{array}{l}
\left| {{W_t}} \right\rangle  = {t^{ - 1/2}}\left( {{{\left| 1 \right\rangle }_1}{{\left| 0 \right\rangle }_2} \ldots {{\left| 0 \right\rangle }_t} + {{\left| 0 \right\rangle }_1}{{\left| 1 \right\rangle }_2} \ldots {{\left| 0 \right\rangle }_t}} {+  \ldots  + {{\left| 0 \right\rangle }_1}{{\left| 0 \right\rangle }_2} \ldots {{\left| 1 \right\rangle }_t}} \right)
\end{array}
\end{equation}

In order to describe it more clearly, we take ${\left| {{W_1}} \right\rangle _4}{\rm{ = }}{1 \over 2}\left( {\left| {1100} \right\rangle {\rm{ + }}\left| {0110}
\right\rangle {\rm{ + }}\left| {0011} \right\rangle  + \left| {1001} \right\rangle } \right)$ as example to demonstrate as bellow. Its multiplicative MGP subgroup is:

$\begin{array}{l}{G_2}\left( {16} \right) = {G_1} \otimes {G_1}\\
\;\;\;\;\;\;\;\; \;\;\; \;\;\;\;  = \left\{ {I \otimes I,I \otimes X,I \otimes Y,I \otimes Z,X \otimes I,X \otimes X,X \otimes Y,X \otimes Z,}\right.\\
\;\;\;\;\;\;\;\;\;\;\;\;\;\;\; \;\;\;\; \left.{Y \otimes I,Y \otimes X,Y \otimes Y,Y \otimes Z,Z \otimes I,Z \otimes X,Z \otimes Y,Z \otimes Z} \right\}.
\end{array}$\\
Obviously, ${G_2}\left( {16} \right)$ directly satisfies Condition 1. And referring to Condition 2,
the positions of the two operated qubits are arbitrary, Tab.~\ref{tab4} illustrates
that ${G_2}\left( {16} \right)$ can be operated on qubit 1 and 2 of ${\left| {{W_1}} \right\rangle _4}$ to implement maximal dense coding.

\begin{table*}[htbp]
\caption{Maximal dense coding of ${\left| {{W_1}} \right\rangle _4}$ using the elements of ${G_2}\left( {16} \right)$}
\label{tab4}
\tabcolsep 15pt 
\begin{tabular}{ll}
\hline\noalign{\smallskip}
Unitary operators on qubits 1 and 2 & 4-qubit W state ${\left| {{W_1}} \right\rangle _4}$\\
\noalign{\smallskip}\hline\noalign{\smallskip}
  ${U_0}$=I$ \otimes $I &
  $\frac{1}{2}{\rm{(\left|{1100} \right\rangle+\left|{0110} \right\rangle+\left|{0011} \right\rangle+\left|{1001} \right\rangle)}}$\\
  ${U_1}$=I$ \otimes $X & $\frac{1}{2}{\rm{(\left|{1000} \right\rangle+\left|{0010} \right\rangle+\left|{0111} \right\rangle+\left|{1101} \right\rangle)}}$\\
  ${U_2}$=I$ \otimes $Y & $\frac{1}{2}{\rm{(-\left|{1100} \right\rangle-\left|{0110} \right\rangle+\left|{0011} \right\rangle+\left|{1001} \right\rangle)}}$\\
  ${U_3}$=I$ \otimes $Z & $\frac{1}{2}{\rm{(\left|{1000} \right\rangle+\left|{0010} \right\rangle-\left|{0111} \right\rangle-\left|{1101} \right\rangle)}}$\\
  ${U_4}$=X$ \otimes $I & $\frac{1}{2}{\rm{(\left|{0100} \right\rangle+\left|{1110} \right\rangle+\left|{1011} \right\rangle+\left|{0001} \right\rangle)}}$\\
  ${U_5}$=X$ \otimes $X & $\frac{1}{2}{\rm{(\left|{0000} \right\rangle+\left|{1010} \right\rangle+\left|{1111} \right\rangle+\left|{0101} \right\rangle)}}$\\
  ${U_6}$=X$ \otimes $Y & $\frac{1}{2}{\rm{(\left|{0000} \right\rangle+\left|{1010} \right\rangle-\left|{1111} \right\rangle-\left|{0101} \right\rangle)}}$\\
  ${U_7}$=X$ \otimes $Z & $\frac{1}{2}{\rm{(-\left|{0100} \right\rangle-\left|{1110} \right\rangle+\left|{1011} \right\rangle+\left|{0001} \right\rangle)}}$\\
  ${U_8}$=Y$ \otimes $I & $\frac{1}{2}{\rm{(\left|{0000} \right\rangle-\left|{1010} \right\rangle-\left|{1111} \right\rangle+\left|{0101} \right\rangle)}}$\\
  ${U_9}$=Y$ \otimes $X & $\frac{1}{2}{\rm{(\left|{0100} \right\rangle-\left|{1110} \right\rangle-\left|{1011} \right\rangle+\left|{0001} \right\rangle)}}$\\
  ${U_{10}}$=Y$ \otimes $Y & $\frac{1}{2}{\rm{(-\left|{0100} \right\rangle+\left|{1110} \right\rangle-\left|{1011} \right\rangle+\left|{0001} \right\rangle)}}$\\
  ${U_{11}}$=Y$ \otimes $Z & $\frac{1}{2}{\rm{(\left|{0000} \right\rangle-\left|{1010} \right\rangle+\left|{1111} \right\rangle-\left|{0101} \right\rangle)}}$\\
  ${U_{12}}$=Z$ \otimes $I & $\frac{1}{2}{\rm{(-\left|{1100} \right\rangle+\left|{0110} \right\rangle+\left|{0011} \right\rangle-\left|{1001} \right\rangle)}}$\\
  ${U_{13}}$=Z$ \otimes $X & $\frac{1}{2}{\rm{(-\left|{1000} \right\rangle+\left|{0010} \right\rangle+\left|{0111} \right\rangle-\left|{1101} \right\rangle)}}$\\
  ${U_{14}}$=Z$ \otimes $Y & $\frac{1}{2}{\rm{(-\left|{1000} \right\rangle+\left|{0010} \right\rangle-\left|{0111} \right\rangle+\left|{1101} \right\rangle)}}$\\
  ${U_{15}}$=Z$ \otimes $Z & $\frac{1}{2}{\rm{(\left|{1100} \right\rangle-\left|{0110} \right\rangle+\left|{0011} \right\rangle-\left|{1001} \right\rangle)}}$\\
\noalign{\smallskip}\hline
\end{tabular}
\end{table*}

\textbf{(3) 4-qubit and 5-qubit cluster states}

In quantum information and quantum computing, cluster state ${\left| {{\phi _{\left\{ k \right\}}}} \right\rangle _C}$
is relative common non-maximal entanglement state which obey the set eigenvalue equations~\cite{Briegel2001Persistent}:
\begin{equation}
\label{eqn:09}
{K^{\left( a \right)}}{\left| {{\phi _{\left\{ k \right\}}}} \right\rangle _C} = {\left( { - 1} \right)^{{k_a}}}{
\left| {{\phi _{\left\{ k \right\}}}} \right\rangle _C},
\end{equation}
where ${K^{\left( a \right)}}$ are the correlation operators
\begin{equation}
\label{eqn:10}
{K^{\left( a \right)}} = \sigma _x^{\left( a \right)}\mathop
 \otimes \limits_{b \in N\left( a \right)} \sigma _z^{\left( b
  \right)},
\end{equation}
\\here, ${\sigma _x}$ and ${\sigma _z}$ are Pauli matrices, $N\left( a \right)$ denotes the neighbourhood of $a$ and $\left\{ {{k_a} \in \left\{ {0,1} \right\}\left| {a \in C} \right.} \right\}$ is a set of binary parameters specifying the particular instance of a cluster state.

According to Eq.~\ref{eqn:09}, it is apparent that one of the 4-qubit, 5-qubit cluster states
can be expressed as
$\left|{c} \right\rangle_4 = \frac{1}{2}{\left( {\left|{0000 } \right\rangle  + \left|{0011 } \right\rangle  +
\left|{1100 } \right\rangle - \left|{1111 } \right\rangle } \right)_{1234}}$ and
$\left|{c} \right\rangle_5 {\rm{ = }}\frac{1}{2}{\left( {\left|{00000 } \right\rangle {\rm{ + }}\left|{00111 }
\right\rangle {\rm{ + }}\left|{11011 } \right\rangle {\rm{ - }}
\left|{11100 } \right\rangle } \right)_{12345}}$, respectively. Taking the above two states for example, we show
how to seek the appropriate unitary operator sets of cluster state.

For 4-qubit cluster state, its multiplicative MGP subgroup is as below:

$\begin{array}{l}{G_2}\left( {16} \right) = {G_1} \otimes {G_1}\\
\;\;\;\;\;\;\;\;\;\;\;\;\;\;\; = \left\{ {I \otimes I,I \otimes X,I \otimes Y,I \otimes Z,X \otimes I,X \otimes X,X \otimes Y,X \otimes Z,}\right.\\
\;\;\;\;\;\;\;\;\;\;\;\;\;\;\;\;\;\;\;\left.{Y \otimes I,Y \otimes X,Y \otimes Y,Y \otimes Z,Z \otimes I,Z \otimes X,Z \otimes Y,Z \otimes Z} \right\}.
\end{array}$\\
Obviously, ${G_2}\left( {16} \right)$ directly satisfies Condition 1. For ${\left| c \right\rangle _4}$, Condition 2 has a certain constraint on the position of corresponding operated qubits, that is, one of the operated qubits must
be chosen from qubit 1 or 2, and the other is qubit 3 or 4.
Tab.~\ref{tab1} illustrates
that ${G_2}\left( {16} \right)$ can be operated on qubit 1 and 4 of 4-qubit cluster state for maximal dense coding.

\begin{table*}[htbp]
\caption{Maximal dense coding of 4-qubit cluster state using the elements of ${G_2}\left( {16} \right)$}
\label{tab1}
\tabcolsep 15pt 
\begin{tabular}{ll}
\hline\noalign{\smallskip}
Unitary operators on qubits 1 and 4 & 4-qubit cluster state\\
\noalign{\smallskip}\hline\noalign{\smallskip}
  ${U_0}$=I$ \otimes $I &
  $\frac{1}{2}{\rm{(\left|{0000} \right\rangle+\left|{0011} \right\rangle+\left|{1100} \right\rangle-\left|{1111} \right\rangle)}}$\\
  ${U_1}$=I$ \otimes $X & $\frac{1}{2}{\rm{(\left|{0001} \right\rangle+\left|{0010} \right\rangle+\left|{1101} \right\rangle-\left|{1110} \right\rangle)}}$\\
  ${U_2}$=I$ \otimes $Y & $\frac{1}{2}{\rm{(-\left|{0001} \right\rangle+\left|{0010} \right\rangle-\left|{1101} \right\rangle-\left|{1110} \right\rangle)}}$\\
  ${U_3}$=I$ \otimes $Z & $\frac{1}{2}{\rm{(\left|{0000} \right\rangle-\left|{0011} \right\rangle+\left|{1100} \right\rangle+\left|{1111} \right\rangle)}}$\\
  ${U_4}$=X$ \otimes $I & $\frac{1}{2}{\rm{(\left|{1000} \right\rangle+\left|{1011} \right\rangle+\left|{0100} \right\rangle-\left|{0111} \right\rangle)}}$\\
  ${U_5}$=X$ \otimes $X & $\frac{1}{2}{\rm{(\left|{1001} \right\rangle+\left|{1010} \right\rangle+\left|{0101} \right\rangle-\left|{0110} \right\rangle)}}$\\
  ${U_6}$=X$ \otimes $Y & $\frac{1}{2}{\rm{(-\left|{1001} \right\rangle+\left|{1010} \right\rangle-\left|{0101} \right\rangle-\left|{0110} \right\rangle)}}$\\
  ${U_7}$=X$ \otimes $Z & $\frac{1}{2}{\rm{(\left|{1000} \right\rangle-\left|{1011} \right\rangle+\left|{0100} \right\rangle+\left|{0111} \right\rangle)}}$\\
  ${U_8}$=Y$ \otimes $I & $\frac{1}{2}{\rm{(-\left|{1000} \right\rangle-\left|{1011} \right\rangle+\left|{0100} \right\rangle-\left|{0111} \right\rangle)}}$\\
  ${U_9}$=Y$ \otimes $X & $\frac{1}{2}{\rm{(-\left|{1001} \right\rangle-\left|{1010} \right\rangle+\left|{0101} \right\rangle-\left|{0110} \right\rangle)}}$\\
  ${U_{10}}$=Y$ \otimes $Y & $\frac{1}{2}{\rm{(\left|{1001} \right\rangle-\left|{1010} \right\rangle-\left|{0101} \right\rangle-\left|{0110} \right\rangle)}}$\\
  ${U_{11}}$=Y$ \otimes $Z & $\frac{1}{2}{\rm{(-\left|{1000} \right\rangle+\left|{1011} \right\rangle+\left|{0100} \right\rangle+\left|{0111} \right\rangle)}}$\\
  ${U_{12}}$=Z$ \otimes $I & $\frac{1}{2}{\rm{(\left|{0000} \right\rangle+\left|{0011} \right\rangle-\left|{1100} \right\rangle+\left|{1111} \right\rangle)}}$\\
  ${U_{13}}$=Z$ \otimes $X & $\frac{1}{2}{\rm{(\left|{0001} \right\rangle+\left|{0010} \right\rangle-\left|{1101} \right\rangle+\left|{1110} \right\rangle)}}$\\
  ${U_{14}}$=Z$ \otimes $Y & $\frac{1}{2}{\rm{(-\left|{0001} \right\rangle+\left|{0010} \right\rangle+\left|{1101} \right\rangle+\left|{1110} \right\rangle)}}$\\
  ${U_{15}}$=Z$ \otimes $Z & $\frac{1}{2}{\rm{(\left|{0000} \right\rangle-\left|{0011} \right\rangle-\left|{1100} \right\rangle-\left|{1111} \right\rangle)}}$\\
\noalign{\smallskip}\hline
\end{tabular}
\end{table*}

For 5-qubit cluster state,
the multiplicative MGP subgroups are as follows,

$G_3^1\left( {32} \right){\rm{ = }}G_2^1 \otimes {G_1}$

$G_3^2\left( {32} \right){\rm{ = }}G_2^2 \otimes {G_1}$

$G_3^3\left( {32} \right){\rm{ = }}G_2^3 \otimes {G_1}$

$G_3^4\left( {32} \right){\rm{ = }}G_2^4 \otimes {G_1}$

$G_3^5\left( {32} \right){\rm{ = }}G_2^5 \otimes {G_1}$

$G_3^6\left( {32} \right){\rm{ = }}G_2^6 \otimes {G_1}$

$G_3^7\left( {32} \right){\rm{ = }}G_2^7 \otimes {G_1}$

$G_3^8\left( {32} \right){\rm{ = }}G_2^8 \otimes {G_1}$

$G_3^9\left( {32} \right){\rm{ = }}G_2^9 \otimes {G_1}$

$G_3^{10}\left( {32} \right){\rm{ = }}G_2^{10} \otimes {G_1}$

$G_3^{11}\left( {32} \right){\rm{ = }}G_2^{11} \otimes {G_1}$

$G_3^{12}\left( {32} \right){\rm{ = }}G_2^{12} \otimes {G_1}$

$G_3^{13}\left( {32} \right){\rm{ = }}G_2^{13} \otimes {G_1}$

$G_3^{14}\left( {32} \right){\rm{ = }}G_2^{14} \otimes {G_1}$

$G_3^{15}\left( {32} \right){\rm{ = }}G_2^{15} \otimes {G_1}$.
\\Referring to Condition 1, we know the effect of $I \otimes I \otimes I$ is the same as $Z \otimes Z \otimes I$,
so $G_3^1,G_3^2,G_3^3,G_3^6,G_3^7,G_3^8,G_3^9,G_3^{10},G_3^{11},G_3^{13},G_3^{15}$ need to be excluded. That is, the appropriate unitary operator sets are $G_3^4,G_3^5,G_3^{12},G_3^{14}$.
If we swap the first column and the second column of $G_3^4,G_3^5,G_3^{12},G_3^{14}$,
the other four unitary operator sets (i.e., $G_3^{4'},G_3^{5'},G_3^{12'},G_3^{14'}$) are obtained.
And referring to Condition 2, the positions of the three operated qubits are arbitrary.
The appropriate unitary operator sets are shown concretely as follows,

$\begin{array}{l}
G_3^4\left( {32} \right){\rm{ = }}\left( {\left\{ {I,X} \right\} \otimes \left\{ {I,Y} \right\} \otimes {G_1}} \right)\cup \left( {\left\{ {I,X} \right\} \otimes \left\{ {X,Z} \right\} \otimes {G_1}} \right)
\end{array}$

$\begin{array}{l}
G_3^5\left( {32} \right){\rm{ = }}\left( {\left\{ {I,Y} \right\} \otimes \left\{ {I,Y} \right\} \otimes {G_1}} \right)\cup \left( {\left\{ {I,Y} \right\} \otimes \left\{ {X,Z} \right\} \otimes {G_1}} \right)
\end{array}$

$\begin{array}{l}
G_3^{12}\left( {32} \right){\rm{ = }}\left( {\left\{ {I,X} \right\} \otimes \left\{ {I,Z} \right\} \otimes {G_1}} \right)\cup \left( {\left\{ {Y,Z} \right\} \otimes \left\{ {Y,X} \right\} \otimes {G_1}} \right)
\end{array}$

$\begin{array}{l}
G_3^{14}\left( {32} \right){\rm{ = }}\left( {\left\{ {I,Y} \right\} \otimes \left\{ {I,Z} \right\} \otimes {G_1}} \right)\cup \left( {\left\{ {X,Z} \right\} \otimes \left\{ {Y,X} \right\} \otimes {G_1}} \right)
\end{array}$

$\begin{array}{l}
G_3^{4'}\left( {32} \right){\rm{ = }}\left( {\left\{ {I,Y} \right\} \otimes \left\{ {I,X} \right\} \otimes {G_1}} \right)\cup \left( {\left\{ {X,Z} \right\} \otimes \left\{ {I,X} \right\} \otimes {G_1}} \right)
\end{array}$

$\begin{array}{l}
G_3^{5'}\left( {32} \right){\rm{ = }}\left( {\left\{ {I,Y} \right\} \otimes \left\{ {I,Y} \right\} \otimes {G_1}} \right)\cup \left( {\left\{ {X,Z} \right\} \otimes \left\{ {I,Y} \right\} \otimes {G_1}} \right)
\end{array}$

$\begin{array}{l}
G_3^{12'}\left( {32} \right){\rm{ = }}\left( {\left\{ {I,Z} \right\} \otimes \left\{ {I,X} \right\} \otimes {G_1}} \right)\cup \left( {\left\{ {Y,X} \right\} \otimes \left\{ {Y,Z} \right\} \otimes {G_1}} \right)
\end{array}$

$\begin{array}{l}
G_3^{14'}\left( {32} \right){\rm{ = }}\left( {\left\{ {I,Z} \right\} \otimes \left\{ {I,Y} \right\} \otimes {G_1}} \right)\cup \left( {\left\{ {Y,X} \right\} \otimes \left\{ {X,Z} \right\} \otimes {G_1}} \right)
\end{array}$.
\\here, $G_3^4,G_3^5,G_3^{4'},G_3^{5'}$ belong to the case of $s = 2$, and $G_3^{12},G_3^{14},G_3^{12'},G_3^{14'}$ belong
 to the case of $s = 1$, and maximal dense coding of 5-qubit cluster state using $G_3^{4'}\left( {32} \right)$ is demonstrated in Tab.~\ref{tab2}.

\begin{table*}[htbp]
\caption{Maximal dense coding of 5-qubit cluster state using the elements of $G_3^{4'}\left( {32} \right)$}
\label{tab2}
\tabcolsep 15pt 
\begin{tabular}{ll}
\hline\noalign{\smallskip}
Unitary operators on qubits 1, 2and 3 & 5-qubit cluster state\\
\noalign{\smallskip}\hline\noalign{\smallskip}
  ${U_0}$=I$ \otimes $I$ \otimes $I & $\frac{1}{2}{\rm{(\left|{00000} \right\rangle+\left|{00111} \right\rangle+\left|{11011} \right\rangle-\left|{11100} \right\rangle)}}$\\
  ${U_1}$=I$ \otimes $X$ \otimes $I & $\frac{1}{2}{\rm{(\left|{01000} \right\rangle+\left|{01111} \right\rangle+\left|{10011} \right\rangle-\left|{10100} \right\rangle)}}$\\
  ${U_2}$=Y$ \otimes $I$ \otimes $I & $\frac{1}{2}{\rm{(-\left|{10000} \right\rangle-\left|{10111} \right\rangle+\left|{01011} \right\rangle-\left|{01100} \right\rangle)}}$\\
  ${U_3}$=Y$ \otimes $X$ \otimes $I & $\frac{1}{2}{\rm{(-\left|{11000} \right\rangle-\left|{11111} \right\rangle+\left|{00011} \right\rangle-\left|{00100} \right\rangle)}}$\\
  ${U_4}$=I$ \otimes $I$ \otimes $X & $\frac{1}{2}{\rm{(\left|{00100} \right\rangle+\left|{00011} \right\rangle+\left|{11111} \right\rangle-\left|{11000} \right\rangle)}}$\\
  ${U_5}$=I$ \otimes $X$ \otimes $X & $\frac{1}{2}{\rm{(\left|{01100} \right\rangle+\left|{01011} \right\rangle+\left|{10111} \right\rangle-\left|{10000} \right\rangle)}}$\\
  ${U_6}$=Y$ \otimes $I$ \otimes $X & $\frac{1}{2}{\rm{(-\left|{10100} \right\rangle-\left|{10011} \right\rangle+\left|{01111} \right\rangle-\left|{01000} \right\rangle)}}$\\
  ${U_7}$=Y$ \otimes $X$ \otimes $X & $\frac{1}{2}{\rm{(-\left|{11100} \right\rangle-\left|{11011} \right\rangle+\left|{00111} \right\rangle-\left|{00000} \right\rangle)}}$\\
  ${U_8}$=I$ \otimes $I$ \otimes $Y & $\frac{1}{2}{\rm{(-\left|{00100} \right\rangle+\left|{00011} \right\rangle-\left|{11111} \right\rangle-\left|{11000} \right\rangle)}}$\\
  ${U_9}$=I$ \otimes $X$ \otimes $Y & $\frac{1}{2}{\rm{(-\left|{01100} \right\rangle+\left|{01011} \right\rangle-\left|{10111} \right\rangle-\left|{10000} \right\rangle)}}$\\
  ${U_{10}}$=Y$ \otimes $I$ \otimes $Y & $\frac{1}{2}{\rm{(\left|{10100} \right\rangle-\left|{10011} \right\rangle-\left|{01111} \right\rangle-\left|{01000} \right\rangle)}}$\\
  ${U_{11}}$=Y$ \otimes $X$ \otimes $Y & $\frac{1}{2}{\rm{(\left|{11100} \right\rangle-\left|{11011} \right\rangle-\left|{00111} \right\rangle-\left|{00000} \right\rangle)}}$\\
  ${U_{12}}$=I$ \otimes $I$ \otimes $Z & $\frac{1}{2}{\rm{(\left|{00000} \right\rangle-\left|{00111} \right\rangle+\left|{11011} \right\rangle+\left|{11100} \right\rangle)}}$\\
  ${U_{13}}$=I$ \otimes $X$ \otimes $Z& $\frac{1}{2}{\rm{(\left|{01000} \right\rangle-\left|{01111} \right\rangle+\left|{10011} \right\rangle+\left|{10100} \right\rangle)}}$\\
  ${U_{14}}$=Y$ \otimes $I$ \otimes $Z & $\frac{1}{2}{\rm{(-\left|{10000} \right\rangle+\left|{10111} \right\rangle+\left|{01011} \right\rangle+\left|{01100} \right\rangle)}}$\\
  ${U_{15}}$=Y$ \otimes $X$ \otimes $Z & $\frac{1}{2}{\rm{(-\left|{11000} \right\rangle+\left|{11111} \right\rangle+\left|{00011} \right\rangle+\left|{00100} \right\rangle)}}$\\
  ${U_{16}}$=I$ \otimes $Y$ \otimes $I & $\frac{1}{2}{\rm{(-\left|{01000} \right\rangle-\left|{01111} \right\rangle+\left|{10011} \right\rangle-\left|{10100} \right\rangle)}}$\\
  ${U_{17}}$=I$ \otimes $Z$ \otimes $I & $\frac{1}{2}{\rm{(\left|{00000} \right\rangle+\left|{00111} \right\rangle-\left|{11011} \right\rangle+\left|{11100} \right\rangle)}}$\\
  ${U_{18}}$=Y$ \otimes $Y$ \otimes $I& $\frac{1}{2}{\rm{(\left|{11000} \right\rangle+\left|{11111} \right\rangle+\left|{00011} \right\rangle-\left|{00100} \right\rangle)}}$\\
  ${U_{19}}$=Y$ \otimes $Z$ \otimes $I & $\frac{1}{2}{\rm{(-\left|{10000} \right\rangle-\left|{10111} \right\rangle-\left|{01011} \right\rangle+\left|{01100} \right\rangle)}}$\\
  ${U_{20}}$=I$ \otimes $Y$ \otimes $X & $\frac{1}{2}{\rm{(-\left|{01100} \right\rangle-\left|{01011} \right\rangle+\left|{10111} \right\rangle-\left|{10000} \right\rangle)}}$\\
  ${U_{21}}$=I$ \otimes $Z$ \otimes $X & $\frac{1}{2}{\rm{(\left|{00100} \right\rangle+\left|{00011} \right\rangle-\left|{11111} \right\rangle+\left|{11000} \right\rangle)}}$\\
  ${U_{22}}$=Y$ \otimes $Y$ \otimes $X & $\frac{1}{2}{\rm{(\left|{11100} \right\rangle+\left|{11011} \right\rangle+\left|{00111} \right\rangle-\left|{00000} \right\rangle)}}$\\
  ${U_{23}}$=Y$ \otimes $Z$ \otimes $X & $\frac{1}{2}{\rm{(-\left|{10100} \right\rangle-\left|{10011} \right\rangle-\left|{01111} \right\rangle+\left|{01000} \right\rangle)}}$\\
  ${U_{24}}$=I$ \otimes $Y$ \otimes $Y & $\frac{1}{2}{\rm{(\left|{01100} \right\rangle-\left|{01011} \right\rangle-\left|{10111} \right\rangle-\left|{10000} \right\rangle)}}$\\
  ${U_{25}}$=I$ \otimes $Z$ \otimes $Y & $\frac{1}{2}{\rm{(-\left|{00100} \right\rangle+\left|{00011} \right\rangle+\left|{11111} \right\rangle+\left|{11000} \right\rangle)}}$\\
  ${U_{26}}$=Y$ \otimes $Y$ \otimes $Y & $\frac{1}{2}{\rm{(-\left|{11100} \right\rangle+\left|{11011} \right\rangle-\left|{00111} \right\rangle-\left|{00000} \right\rangle)}}$\\
  ${U_{27}}$=Y$ \otimes $Z$ \otimes $Y & $\frac{1}{2}{\rm{(\left|{10100} \right\rangle-\left|{10011} \right\rangle+\left|{01111} \right\rangle+\left|{01000} \right\rangle)}}$\\
  ${U_{28}}$=I$ \otimes $Y$ \otimes $Z & $\frac{1}{2}{\rm{(-\left|{01000} \right\rangle+\left|{01111} \right\rangle+\left|{10011} \right\rangle+\left|{10100} \right\rangle)}}$\\
  ${U_{29}}$=I$ \otimes $Z$ \otimes $Z & $\frac{1}{2}{\rm{(\left|{00000} \right\rangle-\left|{00111} \right\rangle-\left|{11011} \right\rangle-\left|{11100} \right\rangle)}}$\\
  ${U_{30}}$=Y$ \otimes $Y$ \otimes $Z & $\frac{1}{2}{\rm{(\left|{11000} \right\rangle-\left|{11111} \right\rangle+\left|{00011} \right\rangle+\left|{00100} \right\rangle)}}$\\
  ${U_{31}}$=Y$ \otimes $Z$ \otimes $Z & $\frac{1}{2}{\rm{(-\left|{10000} \right\rangle+\left|{10111} \right\rangle-\left|{01011} \right\rangle-\left|{01100} \right\rangle)}}$\\
\noalign{\smallskip}\hline
\end{tabular}
\end{table*}

\section{Conclusion}

In this study, we proposed a feasible solution of constructing unitary operator sets for quantum maximal dense
coding, which uses minimum qubits to maximally encode a class of $t$-qubit symmetric states. Our solution consists of two
phases: (1) constructing multiplicative MGP subgroups which must be a group under multiplication, and (2)
selecting appropriate unitary operator sets to operate the state. Compared with Shukla et al's method,  we provide detailed steps and conditions which can greatly reduce the construction workload. In addition, our solution can obtain more unitary operator sets for a specific quantum state than Shukla et al's method , which is concluded in Tab.~\ref{tab9}. It should be noted that, for 4-qubit W state, Shukla et al.'s method provides ${G_2^8}\left( {8} \right)$ and ${G_2^9}\left( {8} \right)$ to encode 8 bits message, while our solution provides ${G_2}\left( {16} \right)$ to encode 16 bits message. In some degree, their method is for dense coding and our solution is for maximal dense coding.

Our solution is just suitable for a class of symmetric states, and is not a general solution for every symmetric states. Appendix A gives two examples which violate Constraint 1 or 2. Therefore,
how to find a general unitary operator construction solution is our future direction, and we will try to explore it from the perspective of
entanglement persistence~\cite{Briegel2001Persistent}, translation symmetry~\cite{Cui2012A} and number of superposition items.

\begin{table*}[htbp]
\caption{List of specific quantum states and corresponding unitary operator sets constructed by Shukla et al.'s and our method}
\label{tab9}
\tabcolsep 9pt 
\begin{tabular}{llll}
\hline\noalign{\smallskip}
Quantum state & Shukla et al.'s
& Ours \\
\noalign{\smallskip}\hline\noalign{\smallskip}
  2-qubit Bell state&  ${G_1}\left( {4} \right)$ &${G_1}\left( {4} \right)$\\
  3-qubit GHZ state &  $G_2^1\left( 8 \right),G_2^2\left( 8 \right),G_2^4\left( 8 \right),G_2^5\left( 8 \right)$&
 $G_2^1\left( 8 \right),G_2^2\left( 8 \right),G_2^4\left( 8 \right),G_2^5\left( 8 \right),G_2^{12}\left( 8 \right),G_2^{13}\left( 8 \right),G_2^{14}\left( 8 \right),G_2^{15}\left( 8 \right)$\\
  4-qubit W state&  ${G_2^8}\left( {8} \right)$, ${G_2^9}\left( {8} \right)$ & ${G_2}\left( {16} \right)$ \\
  4-qubit cluster state&  ${G_2}\left( {16} \right)$&${G_2}\left( {16} \right)$\\
  5-qubit cluster state&  $G_3^4\left( {32} \right),G_3^5\left( {32} \right),G_3^7\left( {32} \right),G_3^8\left( {32} \right)$&
  $G_3^4\left( {32} \right),G_3^5\left( {32} \right),G_3^{4'}\left( {32} \right),G_3^{5'}\left( {32} \right),G_3^{12}\left( {32} \right),G_3^{14}\left( {32} \right),G_3^{12'}\left( {32} \right),G_3^{14'}\left( {32} \right)$\\
\noalign{\smallskip}\hline
\end{tabular}
\end{table*}

\begin{acknowledgements}
The authors would like to express heartfelt gratitude to the anonymous reviewers and editor for their comments that improved the quality of this paper. And the support of all the members of the quantum research group of NUIST is especially acknowledged, their professional discussions and advice have helped us a lot. This work was supported by the National Natural Science Foundation of China under Grant 61672290 and 61802002, in part by the Natural Science Foundation of Jiangsu Province under Grant BK20171458, the Natural Science Foundation of the Jiangsu Higher Education Institutions of China(19KJB520028), and the Priority Academic Program Development of Jiangsu Higher Education Institutions(PAPD).
\end{acknowledgements}

%
%


\begin{thebibliography}{}
%
%
\bibitem{Bennett2014Public}
Bennett, C. H., Brassard, G.: Quantum cryptography : public key
distribution and coin tossing. Elsevier B.V. 560, 7-11 (2014)

\bibitem{Paul2013Experimental}
Paul, J., Sebastien, K. J., Anthony, L., Philippe, G., Eleni, D.: Experimental demonstration of long-distance continuous-variable quantum key distribution. Nat. Photonics. 7(5), 378-381 (2013)

\bibitem{Vlachou2018Quantum}
Vlachou, C., Krawec, W., Mateus, P., Paunkovic, N., Souto, A.: Quantum key distribution with quantum walks. Quantum Inf. Process. 17(11), (2018)

\bibitem{Hillery1999Quantum}
Hillery, M., Buzek, V., Berthiaume, A.: Quantum secret sharing.
Phys. Rev. A. 59, 1829-1834 (1999)

\bibitem{Cleve1999How}
Cleve, R., Gottesman, D., Lo, H.: How to share a quantum secret.
Phys. Rev. Lett. 83(3), 648-651 (1999)

\bibitem{Mashhadi2019General}
Mashhadi, S.: General secret sharing based on quantum Fourier transform. Quantum Inf. Process. 18(4), (2019)

\bibitem{Gottesman2000Theory}
Gottesman, D.: Theory of quantum secret sharing. Phys. Rev. A. 61(4), (2000)

\bibitem{Liu2008Efficient}
Liu, W. J., Chen, H. W., Li, Z. Q., Liu, Z. H.: Efficient quantum secure direct communication with authentication. Chinese Physics Letters. 25(7), 2354-2357 (2008)

\bibitem{Yin2018Quantum}
Yin, L. G., Pan, D., Long, G. L.: Quantum Secure Direct Communication: A Survey of Basic Principle and Recent Development. Jurnal Fizik Malaysia. 39(2), 10001-10006 (2018)


\bibitem{Hu2016Experimental}
Hu, J. Y., Yu, B., Jing, M. Y., Xiao, L. T., Jia, S. T., Qin, G. Q., Long, G. L.: Experimental quantum secure direct communication with single photons. Light-Science \& Applications. 5, (2016)

\bibitem{Liu2009An}
Liu, W. J., Chen, H. W., Ma, T. H., Li, Z. Q., Liu, Z. H., Hu, W. B.: An efficient deterministic secure quantum communication scheme based on cluster states and identity authentication. Chinese Physics B, 18(10), 4105-4109 (2009)

\bibitem{Huang2017Efficient}
Huang, W., Su, Q., Liu, B., He, Y. H., Fan, F., Xu, B. J.: Efficient multiparty quantum key agreement with collective detection. Scientific Reports. 7(1), 15264 (2017)

\bibitem{Liu2018AnEfficient}
Liu, W. J., Xu, Y., Yang, C. N., Gao, P. P., Yu, W. B.: An Efficient and Secure Arbitrary N-Party Quantum Key Agreement Protocol Using Bell States. Int. J. Theor. Phys. 57(1), 195-207 (2018)

\bibitem{Lin2019Cryptanalysis}
Lin, S., Guo, G. D., Chen, A. M., Liu, X. F.: Cryptanalysis of multi-party quantum key agreement with five-qubit Brown states. Quantum Inf. Process. 18(12), (2019)

\bibitem{Cao2017Multiparty}
Liu, W. J., Xu, Y., Yang, C. N., Gao, P. P., Yu, W. B.: An Efficient and Secure Arbitrary N-Party Quantum Key Agreement Protocol Using Bell States. Int. J. Theor. Phys. 57(1), 195-207 (2018)

\bibitem{Sheng2017Distributed}
Sheng, Y. B., Zhou, L.: Distributed secure quantum machine learning. Science Bulletin. 62(14), 1025-1029 (2017)

\bibitem{Liu2018Quantum}
Liu, W. J., Gao, P. P., Yu, W. B., Qu, Z. G., Yang, C. N.: Quantum Relief algorithm. Quantum Info. Process. 17(10), 280 (2018)

\bibitem{Biamonte2017Quantum}
Biamonte, J., Wittek, P., Pancotti, N., Rebentrost, P., Wiebe, N., Lloyd, S.: Quantum machine learning. Nature. 549(7671), 195-202 (2017)

\bibitem{Liu2019AUnitary}
Liu, W.; Gao, P.; Wang, Y.; Yu, W.; Zhang, M.: A Unitary Weights Based One-Iteration Quantum Perceptron Algorithm for Non-Ideal Training Sets. IEEE Access. 7, 36854-36865 (2019)


\bibitem{Mattle1996Dense}
Mattle, K., Weinfurter, H., Kwiat, P., Zeilinger, A.: Dense coding in
experimental quantum communication. Phys. Rev. Lett. 76(25), 4656-4659 (1996)

\bibitem{Li2007Complete}
Li, C. Y., Li, X. H., Deng, F. G., Zhou, P., Zhou, H. Y.: Complete multiple round quantum dense coding with quantum logical network. Chinese Science Bulletin. 52(9), 1162-1165 (2007)

\bibitem{Ambainis2002Dense}
Tian, M. B., Zhang, G. F.: Improving the capacity of quantum dense coding by weak measurement and reversal measurement. Quantum Inf. Process. 17(2), (2018)

\bibitem{Hsieh2010Trading}
Hsieh, M. H., Wilde, M. M.: Trading Classical Communication, Quantum Communication, and Entanglement in Quantum Shannon Theory. IEEE Transactions on Information Theory. 56(9), 4705-4730 (2010)

\bibitem{Khalighi2014Survey}
Khalighi, M. A., Uysal, M.: Survey on Free Space Optical Communication: A Communication Theory Perspective. IEEE Communications Surveys and Tutorials. 16(4), 2231-2258 (2014)

\bibitem{Shukla2013On}
Shukla, Chitra, Kothari, Banerjee, Pathak: On the group-theoretic
structure of a class of quantum dialogue;protocols. Phys. Lett. A. 377(7), 518-527 (2013)

\bibitem{Nielson2000Quantum}
Nielson, M. A., Chuang, I. L.: Quantum Computation and Quantum
Information. Cambridge University Press. (2000)


\bibitem{Greenberger1989Going}
Bouwmeester, D., Pan, J.-W., Daniell, M., Weinfurter, H., Zeilinger, A.: Observation of three-photon greenberger-horne-zeilinger entanglement. Phys. Rev. Lett. 82, 1345-1349 (1999)

\bibitem{D2000Three}
Dur, W., Vidal, G., Cirac, J. I.: Three qubits can be entangled in
two inequivalent ways. Phys. Rev. A. 62(6), 062314 (2000)

\bibitem{Briegel2001Persistent}
Briegel, H. J., Raussendorf, R.: Persistent entanglement in arrays of
interacting particles. Phys. Rev. Lett. 86(5), 910-913 (2001)

\bibitem{Cui2012A}
Cui, H. T., Tian, J. L., Wang, C. M., Chen, Y. C.: A classification
of entanglement in multipartite states with translation symmetry. Eur. Phys. J. D. 67(7), 348-354 (2012)
\end{thebibliography}


\section*{Appendix}
\addcontentsline{left}{chapter}{Appendix}
\section*{Our solution is not applicable to the state which violates the two constraints}

In this paper, a feasible solution of constructing unitary operator sets for quantum maximal dense coding is proposed, which uses minimum qubits to encode a class of $t$-qubit symmetric states. These states have two constraints: (1) the number of superposition items of quantum states must be even. (2) there is at least one set of $\left\lceil {{t \over 2}} \right\rceil $ qubits whose superposition items are orthogonal to each other.

That is to say, our solution is not suitable for those quantum states which violate the two constraints. In order to describe it more clearly, we take $\left|{W} \right\rangle_3 {\rm{ = }}\frac{1}{{\sqrt 3 }}
\left( {\left|{001 } \right\rangle {\rm{ + }}\left|{010}\right\rangle {\rm{ + }}\left|{100 } \right\rangle } \right)$
and
${\left| {{W_2}} \right\rangle _4}{\rm{ = }}{1 \over 2}\left( {\left| {0001} \right\rangle {\rm{ + }}\left| {0010}
\right\rangle {\rm{ + }}\left| {0100} \right\rangle  + \left| {1000} \right\rangle } \right)$  as examples to demonstrate as bellow.

(1) $\left|{W} \right\rangle_3 {\rm{ = }}\frac{1}{{\sqrt 3 }}
\left( {\left|{001 } \right\rangle {\rm{ + }}\left|{010}\right\rangle {\rm{ + }}\left|{100 } \right\rangle } \right)$

Obviously, $\left|{W} \right\rangle_3$ violates Constraint 1.
Its multiplicative MGP subgroups after the construction phase are as follows,

$\begin{array}{l}G_2^1\left( 8 \right){\rm{ = }}\left( {\left\{ {I,\left. X \right\} \otimes \left\{ {\left. {I,X} \right\}} \right.} \right.} \right) \cup \left( {\left\{ {Y,\left. Z \right\} \otimes \left\{ {\left. {I,X} \right\}} \right.} \right.} \right)\\
\;\;\;\;\;\;\;\;\;\; =\left\{ {I \otimes I,I \otimes X,X \otimes I,X \otimes X, Y \otimes I,Y \otimes X,Z \otimes I,Z \otimes X} \right\}
\end{array}$

$\begin{array}{l}G_2^2\left( 8 \right){\rm{ = }}\left( {\left\{ {I,\left. X \right\} \otimes \left\{ {\left. {I,Y} \right\}} \right.} \right.} \right) \cup \left( {\left\{ {Y,\left. Z \right\} \otimes \left\{ {\left. {I,Y} \right\}} \right.} \right.} \right)\\
\;\;\;\;\;\;\;\;\;\;=\left\{ {I \otimes I,I \otimes Y,X \otimes I,X \otimes Y,Y \otimes I,Y \otimes Y,Z \otimes I,Z \otimes Y} \right\}
\end{array}$

$\begin{array}{l}G_2^3\left( 8 \right){\rm{ = }}\left( {\left\{ {I,\left. X \right\} \otimes \left\{ {\left. {I,Z} \right\}} \right.} \right.} \right) \cup \left( {\left\{ {Y,\left. Z \right\} \otimes \left\{ {\left. {I,Z} \right\}} \right.} \right.} \right)\\
\;\;\;\;\;\;\;\;\;\;=\left\{ {I \otimes I,I \otimes Z,X \otimes I,X \otimes Z,Y \otimes I,Y \otimes Z,Z \otimes I,Z \otimes Z} \right\}
\end{array}$

$\begin{array}{l}G_2^4\left( 8 \right){\rm{ = }}\left( {\left\{ {I,\left. X \right\} \otimes \left\{ {\left. {I,Y} \right\}} \right.} \right.} \right) \cup \left( {\left\{ {I,\left. X \right\} \otimes \left\{ {\left. {X,Z} \right\}} \right.} \right.} \right)\\
\;\;\;\;\;\;\;\;\;\;=\left\{ {I \otimes I,I \otimes Y,X \otimes I,X \otimes Y,I \otimes X,I \otimes Z,X \otimes X,X \otimes Z} \right\}
\end{array}$

$\begin{array}{l}G_2^5\left( 8 \right){\rm{ = }}\left( {\left\{ {I,\left. {Y} \right\} \otimes \left\{ {\left. {I,Y} \right\}} \right.} \right.} \right) \cup \left( {\left\{ {I,\left. {Y} \right\} \otimes \left\{ {\left. {X,Z} \right\}} \right.} \right.} \right)\\
\;\;\;\;\;\;\;\;\;\;=\left\{ {I \otimes I,I \otimes Y,Y \otimes I,Y \otimes Y,I \otimes X,I \otimes Z,Y \otimes X,Y \otimes Z} \right\}

\end{array}$

$\begin{array}{l}G_2^6\left( 8 \right){\rm{ = }}\left( {\left\{ {I,\left. Z \right\} \otimes \left\{ {\left. {I,Y} \right\}} \right.} \right.} \right) \cup \left( {\left\{ {I,\left. Z \right\} \otimes \left\{ {\left. {X,Z} \right\}} \right.} \right.} \right)\\
\;\;\;\;\;\;\;\;\;\;=\left\{ {I \otimes I,I \otimes Y,Z \otimes I,Z \otimes Y,I \otimes X,I \otimes Z,Z \otimes X,Z \otimes Z} \right\}

\end{array}$

$\begin{array}{l}G_2^7\left( 8 \right){\rm{ = }}\left( {\left\{ {I,\left. Z \right\} \otimes \left\{ {\left. {I,Z} \right\}} \right.} \right.} \right) \cup \left( {\left\{ {X,\left. {Y} \right\} \otimes \left\{ {\left. {X,Y} \right\}} \right.} \right.} \right)\\
\;\;\;\;\;\;\;\;\;\;=\left\{ {I \otimes I,I \otimes Z,Z \otimes I,Z \otimes Z,X \otimes X,X\otimes Y,Y \otimes X,Y \otimes Y} \right\}
\end{array}$

$\begin{array}{l}G_2^8\left( 8 \right){\rm{ = }}\left( {\left\{ {I,\left. {Y} \right\} \otimes \left\{ {\left. {I,X} \right\}} \right.} \right.} \right) \cup \left( {\left\{ {X,\left. Z \right\} \otimes \left\{ {\left. {Z,Y} \right\}} \right.} \right.} \right)\\
\;\;\;\;\;\;\;\;\;\;=\left\{ {I \otimes I,I \otimes X,Y \otimes I,Y \otimes X,X \otimes Z,X \otimes Y,Z \otimes Z,Z \otimes Y} \right\}
\end{array}$

$\begin{array}{l}G_2^9\left( 8 \right){\rm{ = }}\left( {\left\{ {I,\left. X \right\} \otimes \left\{ {\left. {I,Y} \right\}} \right.} \right.} \right) \cup \left( {\left\{ {Z,\left. {Y} \right\} \otimes \left\{ {\left. {Z,X} \right\}} \right.} \right.} \right)\\
\;\;\;\;\;\;\;\;\;\;=\left\{ {I \otimes I,I \otimes Y,X \otimes I,X \otimes Y,Z \otimes Z,Z \otimes X,Y \otimes Z,Y \otimes X} \right\}
\end{array}$

$\begin{array}{l}G_2^{10}\left( 8 \right){\rm{ = }}\left( {\left\{ {I,\left. X \right\} \otimes \left\{ {\left. {I,X} \right\}} \right.} \right.} \right) \cup \left( {\left\{ {Z,\left. {Y} \right\} \otimes \left\{ {\left. {Z,Y} \right\}} \right.} \right.} \right)\\
\;\;\;\;\;\;\;\;\;\;\;\;=\left\{ {I \otimes I,I \otimes X,X \otimes I,X \otimes X,Z \otimes Z,Z \otimes Y,Y \otimes Z,Y \otimes Y} \right\}
\end{array}$

$\begin{array}{l}G_2^{11}\left( 8 \right){\rm{ = }}\left( {\left\{ {I,\left. {Y} \right\} \otimes \left\{ {\left. {I,Y} \right\}} \right.} \right.} \right) \cup \left( {\left\{ {Z,\left. X \right\} \otimes \left\{ {\left. {Z,X} \right\}} \right.} \right.} \right)\\
\;\;\;\;\;\;\;\;\;\;\;\;=\left\{ {I \otimes I,I \otimes Y,Y \otimes I,Y \otimes Y,Z \otimes Z,Z \otimes X,X \otimes Z,X \otimes X} \right\}
\end{array}$

$\begin{array}{l}G_2^{12}\left( 8 \right){\rm{ = }}\left( {\left\{ {I,\left. X \right\} \otimes \left\{ {\left. {I,Z} \right\}} \right.} \right.} \right) \cup \left( {\left\{ {Y,\left. Z \right\} \otimes \left\{ {\left. {Y,X} \right\}} \right.} \right.} \right)\\
\;\;\;\;\;\;\;\;\;\;\;\;=\left\{ {I \otimes I,I \otimes Z,X \otimes I,X \otimes Z,Y \otimes X,Y \otimes Y,Z \otimes X,Z \otimes Y} \right\}
\end{array}$

$\begin{array}{l}G_2^{13}\left( 8 \right){\rm{ = }}\left( {\left\{ {I,\left. Z \right\} \otimes \left\{ {\left. {I,X} \right\}} \right.} \right.} \right) \cup \left( {\left\{ {Y,\left. X \right\} \otimes \left\{ {\left. {Y,Z} \right\}} \right.} \right.} \right)\\
\;\;\;\;\;\;\;\;\;\;\;\;=\left\{ {I \otimes I,I \otimes X,Z \otimes I,Z \otimes X,Y \otimes Y,Y \otimes Z,X \otimes Y,X \otimes Z} \right\}
\end{array}$

$\begin{array}{l}
G_2^{14}\left( 8 \right){\rm{ = }}\left( {\left\{ {I,\left. {Y} \right\} \otimes \left\{ {\left. {I,Z} \right\}} \right.} \right.} \right) \cup \left( {\left\{ {X,\left. Z \right\} \otimes \left\{ {\left. {Y,X} \right\}} \right.} \right.} \right)\\
\;\;\;\;\;\;\;\;\;\;\;\;=\left\{ {I \otimes I,I \otimes Z,{Y} \otimes I,{Y} \otimes Z,Z \otimes X,Z \otimes Y,X \otimes X,X \otimes Y} \right\}
\end{array}$

$\begin{array}{l}
G_2^{15}\left( 8 \right){\rm{ = }}\left( {\left\{ {I,\left. Z \right\} \otimes \left\{ {\left. {I,Y} \right\}} \right.} \right.} \right) \cup \left( {\left\{ {X,\left. {Y} \right\} \otimes \left\{ {\left. {Z,X} \right\}} \right.} \right.} \right)\\
\;\;\;\;\;\;\;\;\;\;\;\;=\left\{ {I \otimes I,I \otimes {Y},{Z} \otimes I,{Z} \otimes {Y},{X} \otimes {Z,X} \otimes {X},{Y} \otimes {Z},{Y} \otimes {X}} \right\}.
\end{array}$\\
Now, we exclude $G_2^3,G_2^6,G_2^7,G_2^8,G_2^9,G_2^{10},G_2^{11}$ who violate Condition 1, that is,
the appropriate unitary operator sets
are $G_2^1,G_2^2,G_2^4,G_2^5,G_2^{12},G_2^{13},G_2^{14},G_2^{15}$. And referring to Condition 2,
the positions of the two operated qubits are arbitrary. Unfortunately,
all sets we constructed could not implement maximal dense coding with minimal qubits. Therefore, our solution is
not suitable for $\left|{W} \right\rangle_3$. Taking $G_2^4\left( 8 \right)$ as an example,
Tab.~\ref{tab3} illustrates the output states of it are not mutually orthogonal which cannot implement maximal dense coding with 2 qubits.

\begin{table}[htbp]
\footnotesize
\centering
\caption{Maximal dense coding of 3-qubit W state(incorrect,just for explanation)}
\label{tab3}
\scalebox{1}[1]{
\tabcolsep 15pt 
\begin{tabular}{ll}
\toprule
\tabincell{c}{ Unitary operators \\on qubits 1 and 2} & 3-qubit W state\\\hline
  ${U_0}$=I$ \otimes $I &
  $\frac{1}{{\sqrt 3 }}{\rm{(\left|{001 }\right\rangle  + \left|{010 }\right\rangle  + \left|{100 }\right\rangle )}}$\\

  ${U_1}$=I$ \otimes $X &
  $\frac{1}{{\sqrt 3 }}\left( {\left| {011 }\right\rangle  +  \left|{000 }\right\rangle  + \left|{110 }\right\rangle } \right)$\\

   ${U_2}$=X$ \otimes $I &
  $\frac{1}{{\sqrt 3 }}\left( {\left| {101 }\right\rangle  + \left|{110 }\right\rangle  + \left|{000 }\right\rangle } \right)$\\

   ${U_3}$=X$ \otimes $X &
  $\frac{1}{{\sqrt 3 }}\left( {\left| {111 }\right\rangle  +  \left|{100 }\right\rangle  + \left|{010 }\right\rangle } \right)$\\

   ${U_4}$=I$ \otimes $Y &
  $\frac{1}{{\sqrt 3 }}\left( { - \left| {011 }\right\rangle  +  \left|{000 }\right\rangle  - \left|{110 }\right\rangle } \right)$\\

   ${U_5}$=I$ \otimes $Z &
  $\frac{1}{{\sqrt 3 }}\left( {\left| {001 }\right\rangle - \left|{010 }\right\rangle  + \left|{100 }\right\rangle } \right)$\\

   ${U_6}$=X$ \otimes $Y &
  $\frac{1}{{\sqrt 3 }}\left( { - \left| {111}\right\rangle  +  \left|{100 }\right\rangle  - \left|{010 }\right\rangle } \right)$\\

   ${U_7}$=X$ \otimes $Z &
  $\frac{1}{{\sqrt 3 }}\left( {\left| {101 }\right\rangle  -  \left|{110 }\right\rangle  + \left|{000 }\right\rangle } \right)$\\
\bottomrule
\end{tabular}
}
\end{table}

(2) ${\left| {{W_2}} \right\rangle _4}{\rm{ = }}{1 \over 2}\left( {\left| {0001} \right\rangle {\rm{ + }}\left| {0010}
\right\rangle {\rm{ + }}\left| {0100} \right\rangle  + \left| {1000} \right\rangle } \right)$.

Obviously, ${\left| {{W_2}} \right\rangle _4}$ violates Constraint 2.
Its multiplicative MGP subgroup is

$\begin{array}{l}{G_2}\left( {16} \right) = {G_1} \otimes {G_1}\\
\;\;\;\;\;\;\;\;\;\;\;\;\;\;\; = \left\{ {I \otimes I,I \otimes X,I \otimes Y,I \otimes Z,X \otimes I,X \otimes X,X \otimes Y,X \otimes Z,}\right.\\
\;\;\;\;\;\;\;\;\;\;\;\;\;\;\;\;\;\;\;\left.{Y \otimes I,Y \otimes X,Y \otimes Y,Y \otimes Z,Z \otimes I,Z \otimes X,Z \otimes Y,Z \otimes Z} \right\}.
\end{array}$\\
Although ${G_2}\left( {16} \right)$ satisfies Condition 1. But, ${\left| {{W_2}} \right\rangle _4}$ does not satisfy Condition 2, that is, we can not select $\left\lceil {{4 \over 2}} \right\rceil {\rm{ = 2}}$ operated qubits which satisfies that the product states of operated qubits are mutually orthogonal. So, ${\left| {{W_2}} \right\rangle _4}$ cannot be used to implement maximal dense coding with 2 qubits.

\end{document}